\documentclass{elsart}

    \usepackage{latexsym,bm,amsmath,amssymb,amsfonts}
    \usepackage{epsfig,graphics,graphicx}
    \usepackage{makeidx}
    \usepackage{citesort}
    \usepackage{slashed}
    \usepackage{float,multicol}
\newcommand{\g}{\gamma}

\newcommand\lr[1]{{\left({#1}\right)}}
\newcommand{\bra}{\langle}
\newcommand{\ket}{\rangle}
\newcommand\ex[1]{{{\rm e}^{#1}}}

    \newcommand{\f}{\frac}
    \newcommand{\rmd}{{\rm d}}   %ELS%
    
    \newcommand{\abar}{\bar{\alpha}_s}

    \newcommand{\lan}{\left\langle}
    \newcommand{\ran}{\right\rangle}
    \newcommand{\cal}{\mathcal}

    \newcommand{\grad}{\nabla}

    \newcommand{\nn}{\nonumber\\}

    \long\def\comment#1{ }
    
    \newcommand{\beq}{\vspace{-.4cm}\begin{eqnarray}}
    \newcommand{\eeq}{\vspace{-.5cm}\end{eqnarray}}
    \newcommand{\be}{\vspace{-.3cm}\begin{eqnarray}}
    \newcommand{\ee}{\vspace{-.4cm}\end{eqnarray}}
\def\simge{\mathrel{%
   \rlap{\raise 0.511ex \hbox{$>$}}{\lower 0.511ex \hbox{$\sim$}}}}
\def\simle{\mathrel{
   \rlap{\raise 0.511ex \hbox{$<$}}{\lower 0.511ex \hbox{$\sim$}}}}

\begin{document}

\begin{flushright}
~\vspace{-1.25cm}\\
{\small\sf SACLAY--T06/044}
\end{flushright}
\vspace{2.cm}

\begin{frontmatter}

\parbox[]{16.0cm}{ \begin{center}
\title{Forward gluon production in hadron--hadron \\
 scattering with Pomeron loops}

\author{E.~Iancu\thanksref{th2}},
\author{C.~Marquet},
\author{G.~Soyez\thanksref{th3}}

\address{Service de Physique Th\'eorique, CEA/DSM/SPhT,  Unit\'e de recherche
associ\'ee au CNRS (URA D2306), CEA Saclay,
        F-91191 Gif-sur-Yvette, France}

\thanks[th2]{Membre du Centre National de la Recherche Scientifique
(CNRS), France.}
\thanks[th3]{On leave from the Fundamental Theoretical Physics group
of the University of Li\`ege.}

\date{\today}
%\vspace{0.8cm}
\begin{abstract}
We discuss new physical phenomena expected in particle production in
hadron--hadron collisions at high energy, as a consequence of
Pomeron loop effects in the evolution equations for the Color Glass
Condensate. We focus on gluon production in asymmetric,
`dilute--dense', collisions : a dilute projectile scatters off a
dense hadronic target, whose gluon distribution is highly evolved.
This situation is representative for particle production in
proton--proton collisions at forward rapidities (say, at LHC) and
admits a dipole factorization similar to that of deep inelastic
scattering (DIS). We show that at sufficiently large forward
rapidities, where the Pomeron loop effects become important in the
evolution of the target wavefunction, gluon production is dominated
by `black spots' (saturated gluon configurations) up to very large
values of the transverse momentum, well above the {\em average}
saturation momentum in the target. In this regime, the produced
gluon spectrum exhibits {\em diffusive scaling}, so like DIS at
sufficiently high energy.

\end{abstract}
\end{center}}

\end{frontmatter}

%\newpage
%\tableofcontents
\newpage

\section{Introduction}
\setcounter{equation}{0} \label{SECT_INTRO}

The recent data on particle production at semi--hard transverse
momenta ($p_\perp= 1\div 5$ GeV)  in deuteron--gold (d+Au)
collisions at RHIC \cite{RHIC-dAu-mid,Brahms-data,STAR-data} show
%the Relativistic Heavy Ion Collider (RHIC) show
strong evidence in favour of high gluon density effects in the
nuclear target and can be naturally accommodated
\cite{KLM02,JNV,KKT,Baier03,Nestor03,BGV04,IIT04,KKT2,BMTS06,Dumitru1,Dumitru2}
within the same theoretical framework that has been previously used
to describe the HERA data at small Bjorken--$x$ \cite{IIM03} : the
effective theory for non--linear evolution in QCD at high energy
known as the ``Color Glass Condensate'' (CGC) \cite{MV,CGC} (see the
review papers \cite{EdiCGC} for details and more references).

In particular, the results at RHIC show a striking difference
between the particle yield at central and, respectively, forward
rapidities
%(with respect to the deuteron fragmentation region)
--- namely, a very rapid evolution of the nuclear modification
factor $R_{\rm dAu}$ from a `Cronin peak' ($R_{\rm dAu}>1$) at
central rapidity ($\eta=0$) to `high--$p_\perp$ suppression'
($R_{\rm dAu} < 1$) at forward rapidity ($\eta > 1$); see, e.g., the
discussion in \cite{Brahms-Review}
--- which is naturally attributed to the small--$x$ evolution of the
gluon distribution in the nucleus in the presence of non--linear
effects \cite{KLM02,KKT,Nestor03,IIT04,Dumitru1}. (One should recall
at this point that increasing $\eta$ is tantamount to probing gluons
with smaller values of $x$ in the nuclear wavefunction; see the
discussion of kinematics in Sect. \ref{SECT_FACT}.) Moreover, for
very forward rapidities ($\eta\ge 3$), the spectrum of the produced
hadrons appears to be consistent with {\em geometric scaling}
\cite{geometric}, which is a hallmark of the BFKL evolution in the
vicinity of the saturation line \cite{SCALING,MT02,DT02}. For
smaller rapidities $0\le\eta < 3$, the {\em violations} of this
scaling become visible too \cite{Dumitru2}, and they indeed follow
the pattern predicted by the CGC effective theory
\cite{SCALING,MT02,DT02} --- a pattern which has been also verified
in the HERA data for the $F_2$ structure function at $x < 0.01$
\cite{IIM03}.

These results at RHIC demonstrate the potential of the
`dilute--dense' (here, deuteron--gold) hadron--hadron collisions to
accurately probe the universal phase of QCD at high energy --- the
Color Glass Condensate. We expect this potential to be amplified in
the incoming experiments at LHC, because of the considerably higher
energies that will be available there. One should emphasize here
that such asymmetric collisions, in which one of the partaking
hadrons (the `projectile') is dilute and the other one (the
`target') is dense, are exceptional in several respects:

$\bullet$ On the {\em experimental} side, the `dilute--dense'
collisions are less affected by the final--state interactions than
the `dense--dense' collisions (like the nucleus--nucleus ones),
hence they represent a relatively clean measurement of the initial
conditions (nearly on the same footing as the deep inelastic
lepton--hadron scattering).

$\bullet$  On the side of the {\em  theory}, such collisions are
better under control, in the sense that factorization schemes are
available, which allow one to compute the cross--section for
particle production as a convolution of (standard) parton
distributions for the dilute projectile times `generalized parton
distributions' (including high--density effects) for the dense
target times partonic cross--sections
\cite{KM98,KTS99,DM01,KT02,KW02,BGV04,CM04}. This factorization is
very similar to the `dipole factorization' used for DIS at high
energy (see, e.g., Ref. \cite{EdiCGC} and references therein), and
the `generalized parton distributions' alluded to above are in fact
{\em dipole--hadron scattering amplitudes} \cite{BraunG}, which
describe the multiple scattering between a `color dipole' and the
dense hadronic target.

\comment{ Whereas in the case of DIS, a `physical' dipole appears
naturally in the problem, as the quark--antiquark fluctuation of the
virtual photon (which then scatters off the hadron), in the case of
dilute--dense hadron--hadron collisions, the `dipole' is merely a
mathematical device which appears when constructing the
cross--section: For instance, if the scattering involves a quark out
of the projectile, then the `dipole' is built at the level of the
cross--section, as the product of the quark line in the direct
amplitude times the corresponding `antiquark' line in the complex
conjugate amplitude (see Sect. \ref{SECT_FACT}); we then speak about
a quark--antiquark ($q\bar q$) dipole. If, on the other hand, the
projectile parton which initiates the scattering is a gluon, we
rather need to consider a gluon--gluon ($gg$) dipole.}

Hence, the non--trivial scattering problem that one has to solve is
that of the scattering between an elementary color dipole and the
high--density hadronic target. To that aim, one can rely on the CGC
formalism \cite{EdiCGC}, where the dipole amplitudes obey an
infinite hierarchy of non--linear evolution equations --- the
Balitsky--JIMWLK hierarchy \cite{B,CGC,JKLW,W}. In the limit where
the number of colors $N_c$ is large, this hierarchy boils down to a
single equation, the Balitsky--Kovchegov (BK) equation \cite{B,K}.
Most of the previous analyses of the d+Au collisions within the
framework of CGC were based on the solution to BK equation, or on
various approximations to it \cite{KLM02,KKT,Nestor03,IIT04}; as
mentioned before, they led to a rather satisfactory picture, which
provides a qualitative explanation\footnote{Quantitative fits have
been also obtained \cite{KKT2,Dumitru1,Dumitru2}, but only at the
expense of introducing some free parameters, to account for the lack
of accuracy of the current formalism and for the uncertainties
associated with the non--perturbative initial conditions. It is
interesting to note that most of these parameters were in fact fixed
from fits to the HERA data for $F_2$.} of the experimental situation
at RHIC and is robust against the various approximations.

But whereas the current CGC formalism, centered around the JIMWLK
equation and its main consequences like `geometric scaling', appears
to be appropriate for understanding the experimental situation at
RHIC and HERA, it is not clear whether this success will extend to
the future data at LHC. Indeed, recent theoretical developments
\cite{IM03,MS04,IMM04,IT04,MSW05,LL05,KL05,BIIT05,BREM,Balit05,HIMST06}
show that some physical ingredients which have been ignored by the
JIMWLK evolution play an increasingly important role with increasing
energy, and should lead to a dramatically different physical picture
at sufficiently high energy. These ingredients are the {\em
gluon--number fluctuations} \cite{AM94,Salam95,IM03,IMM04}
associated with gluon splitting (bremsstrahlung) in the dilute
regime, which act as a seed for $\kappa$--body correlations with
$\kappa\ge 2$. Although formally suppressed by higher powers of the
coupling constant $\alpha_s$, such correlations are rapidly
amplified by the BFKL evolution (the faster the larger is $\kappa$)
and eventually play an essential role in the evolution towards gluon
saturation at high energy. The Balitsky--JIMWLK equations correctly
encode the role of such many--body correlations in providing
saturation, but they fail to include the physical {\em source} for
such correlations, namely, the gluon--number fluctuations at low
density \cite{IT04}. Thus, while these equations are properly
describing the evolution of a dense system towards saturation, they
cannot describe the formation of such a dense system via evolution
from a dilute system at low energy.

At this point, one may be tempted to conclude that the
Balitsky--JIMWLK equations should at least apply to the evolution
of a large nucleus, which starts with a relatively high gluon
density already at low energy. However, this is not true either !
Even for a large nucleus, the gluon distribution involves a dilute
tail at relatively high transverse momenta (well above the nuclear
saturation momentum), and the evolution of that tail is still
dominated by fluctuations. It is nevertheless true that, in order
to probe the effects of fluctuations for such a large nucleus, one
needs to go up to much higher energies, namely, high enough for
the saturation effects to have propagated at relatively large
transverse momenta, which were originally in the dilute tail at
low energy. In that sense, the evolution of a {\em originally
dilute} system, like a proton or a color dipole, is a better
laboratory to study the effects of fluctuations.

The equations describing gluon evolution in the presence of both
saturation and fluctuations are presently known only for large
$N_c$, and are generally referred to as the `Pomeron loop equations'
\cite{IT04,MSW05,LL05,BIIT05}. The `Pomeron' here is the BFKL
Pomeron \cite{BFKL}; namely, this is the amplitude for the
scattering between an external dipole and the target in the linear,
leading--logarithmic, approximation at high energy. In an
appropriate gauge, this amplitude has a diagrammatic interpretation
in terms of ladder diagrams whose rungs are strongly ordered in the
`rapidity' variable $Y\equiv \ln (1/x)$. At large $N_c$, the gluon
bremsstrahlung (responsible for gluon number fluctuations) can be
described as $1 \to 2$ Pomeron splitting, while the saturation
effects are tantamount to $2\to 1$ Pomeron merging. Hence, the
equations describing the complete evolution involve both (Pomeron)
splitting and merging, and thus generate Pomeron loops through
iterations.

The Pomeron loop equations form an infinite hierarchy, with a
complicated non--local structure, and in spite of intense
theoretical efforts, their general solutions are not yet known.
Still, the asymptotic behaviour of the solutions at high energy is
known because of the {\em universality of the stochastic process
described by these equations} \cite{MP03,IMM04,GS05,EGBM05} : this
is in the same universality class as the `reaction--diffusion'
process $A \rightleftharpoons 2A$, which is relevant to a large
variety of problems (in statistical physics, chemistry, biology ...)
and has been intensely scrutinized over the last decades (cite Ref.
\cite{Saar} for a review). But even in that context, it was only
recently realized that this process is extremely sensitive to
particle--number fluctuations in the dilute regime \cite{BD97}. In
the language of QCD, this sensitivity explains why the predictions
of the Pomeron loops equations turn out to be very different from
those of the Balitsky--JIMWLK equations, which ignore fluctuations
\cite{MS04,IMM04,IT04}.

By using known results from statistical physics \cite{BD97,Saar}, it
has been possible to deduce the functional form of the dipole
scattering amplitudes in QCD at high energy and large $N_c$, in
terms of juste a couple of uncontrolled parameters \cite{IT04}. The
most striking feature of these results is the fact that, even in the
{\em weak scattering} regime, that is, for a relatively small
projectile dipole, the scattering amplitude is dominated by {\em
black spots}, i.e., rare gluon configurations which are at
saturation on the resolution scale $Q^2\sim 1/r^2$ of the projectile
($r$ is the dipole size) and thus look `black' to the latter --- the
dipole is completely absorbed when hitting any such a spot. The
reason why the {\em average} amplitude is nevertheless small is
because such `black spots' are relatively rare, so the target disk
looks transparent (or `white') on that resolution scale at most
impact parameters (see Sect. \ref{SECT_DIPOLE} for details).

Very recently, the consequences of this new physical picture have
been explicitly worked out for the case of DIS \cite{HIMST06}, with
the conclusion that new phenomena are to be expected at very high
energy
--- chiefly among them, the replacement of geometric scaling by a
new, {\em diffusive}, scaling which should extend up to very large
$Q^2$ (cf. Sect. \ref{SECT_DIPOLE}). But whereas in the case of DIS,
it is not clear when, and whether, new experiments will become
available with sufficiently high energy to probe this new physics
(as aforementioned, the current experimental situation at HERA is
consistent with geometric scaling, thus suggesting an {\em
intermediate--energy} regime), the imminent advent of the LHC opens
the possibility to observe the Pomeron loops via high--energy
hadron--hadron collisions.

Both $pp$ and $pA$ (and also $AA$) collisions will be performed at
LHC, but for the reasons explained earlier, the best process to
study the fluctuation--dominated regime at high energy is particle
production at forward rapidities in $pp$ collisions. Then, one of
the participating protons (the `projectile') is dilute and acts as a
probe of the small--$x$ part of the wavefunction of the other proton
(the `target'), which is highly evolved. Ideally, one needs very
large forward rapidities for the produced particles, in order to
generate a large asymmetry between the evolutions of the target and
the projectile, and thus probe Pomeron loop effects in the target
proton while keeping the projectile proton dilute.

In view of the theoretical uncertainties alluded to above, it is not
possible for us today to make quantitative predictions for LHC, and
not even to reliably predict whether the new physics should manifest
itself within the kinematical range covered by the LHC, or not. What
we {\em can} do, however, is to predict new, qualitative, phenomena
which would unambiguously signal the new physics in the data for
particle production, and even compute these phenomena in terms of
few free parameters. By comparing these predictions to the data to
become later available at LHC, one should be able to identify this
new physics (if it is present indeed !) and extract the values of
the relevant parameters.

Our main purpose here is to demonstrate the consequences of the
Pomeron loop physics on the particle production in hadron--hadron
collisions at high energy. To that aim, we shall focus on a problem
which is conceptually transparent and avoids unnecessary
complications\footnote{Like the use of realistic parton
distributions or fragmentation functions; note, however, that such
`complications' may become essential and unavoidable in view of
realistic applications to the phenomenology.}: that of the forward
gluon production in onium---hadron scattering at high energy and
large $N_c$.

$\bullet$ The {\em `onium'} is a dilute hadronic system produced via
the BFKL evolution of a small color dipole. Its wavefunction is
explicitly known within perturbative QCD at large $N_c$ \cite{AM94},
and this is why we have chosen it as a projectile. But the
generalization of our subsequent results to a more realistic
projectile, like a (dilute) proton, is straightforward: In the most
interesting kinematical region, the projectile is merely represented
by the standard (integrated) gluon distribution function, as
determined by the leading--twist, DGLAP, approximation.

$\bullet$ The {\em hadronic target} will be assumed to be probed at
very small values of $x$ (corresponding to very forward rapidities
for the produced gluon), in the fluctuation--dominated regime which
is {\em universal\,}: this is the ultimate form of hadronic matter
which is produced at sufficiently high energy starting with
arbitrary initial conditions at low energy. In this regime, the
dipole--target scattering amplitude can be taken in the simplified
functional form predicted by the correspondence with statistical
physics, which represents an approximate solution to the Pomeron
loop equations. This functional form describes not only the
diffusive scaling regime at asymptotically high energies, but also
the geometric scaling regime at intermediate energies, and the
transition between the two. Hence, by increasing the rapidity of the
produced gluon, we shall be able to study the increasing influence
of fluctuations on the cross--section for particle production and
the interplay between geometric and diffusive scaling.

This paper is organized as follows: The first two sections contain
no new results, but rather collect some previous results ---
concerning the dipole factorization of forward gluon production (in
Sect. \ref{SECT_FACT}) and, respectively, the calculation of the
dipole amplitudes in the presence of Pomeron loop (in Sect.
\ref{SECT_DIPOLE}) --- which are necessary for the subsequent
analysis. The first issue (the factorization of the gluon
production) is rather well established in the literature and will be
only succinctly described here. But the second one (the high--energy
evolution with Pomeron loops) has met with important developments
over the last two years, whose results are perhaps less well known
(since spread over several, more technical, papers
\cite{MS04,IMM04,IT04,HIMST06}), yet they are essential for our
present analysis. To cope with that, in Sect. \ref{SECT_DIPOLE} we
shall try to give a self--contained and pedagogical discussion of
these recent results. Then, in Sects. \ref{SECT_GLUON} and
\ref{SECT_SIGMA}, we shall apply them to the analysis of forward
gluon production at high energy. We shall separately discuss the
relevant `generalized gluon distribution' (a special Fourier
transform of the dipole amplitude) (in Sect. \ref{SECT_GLUON}) and
the cross--section for gluon production (in Sect. \ref{SECT_SIGMA}).
We shall point out some interesting differences with respect to the
corresponding analysis of DIS in Ref. \cite{HIMST06}. But in spite
of these differences, the final conclusions are qualitatively
similar: the cross--section for forward gluon production is
dominated by `black spots' up to very large values of the gluon
transverse momentum and the ensuing spectrum shows diffusive
scaling. We shall conclude, in Sect. \ref{SECT_CONC}, with a short
summary and a list of open problems.

\section{Gluon production in onium--hadron scattering: Factorization}
\setcounter{equation}{0}\label{SECT_FACT}

In this section, we shall more precisely describe the calculation
that we intend to perform and recall the factorization scheme which
lies at the basis of this calculation.

As explained in the Introduction, we are interested in gluon
production in the asymmetric collision between a {\em dilute
projectile}, whose wavefunction has been evolved in rapidity up to
$y_1$, and a {\em dense target}, with rapidity $y_2=Y-y_1\gg y_1$.
($Y$ is the total rapidity gap between the projectile and the
target.) Here, by ``dilute'' and ``dense'' we understand that, on
the resolution scale set by the transverse momentum $k_\perp$ of the
produced particle, the gluon distribution in the projectile can be
described in the leading--twist approximation (i.e., its evolution
up to $y_1$ lies within the realm of the linear, DGLAP or BFKL,
evolution equations of QCD), whereas the gluon distribution in the
target is rather the site of important non--linear phenomena, and is
strongly influenced by both saturation and gluon--number
fluctuations --- hence, by Pomeron loops.

\begin{figure}[t]
    \centerline{\epsfxsize=14cm\epsfbox{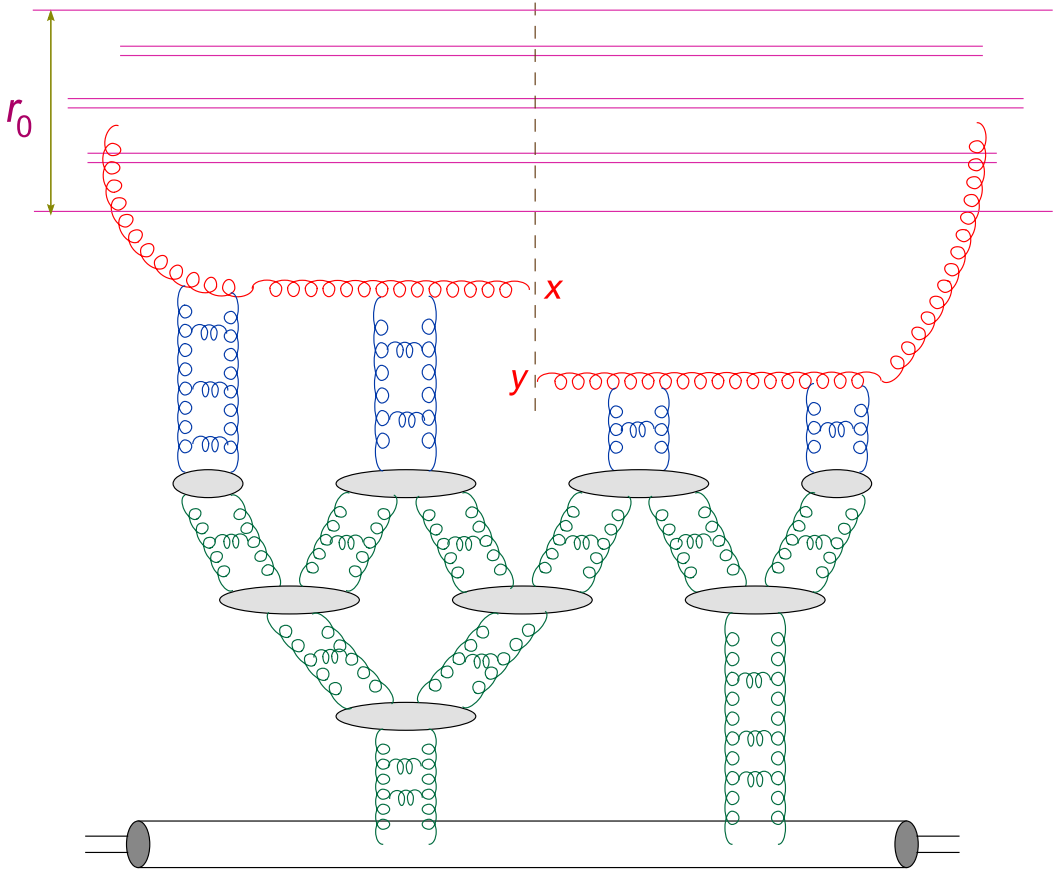}}
    \caption{\sl A schematic illustration of a Feynman graph
    contributing to the cross--section for gluon production
    in onium--hadron scattering in the presence of Pomeron loops.
    The upper part of the graph represents the `onium', i.e., the
    modulus squared of the wavefunction of the BFKL--evolved dipole.
    Each double--line denotes a soft gluon inside the wavefunction
    (the gluon is represented as a pointlike quark--antiquark pair in a
    color octet state, as appropriate at large $N_c$).
    The produced gluon (in red) is radiated by one of the dipoles,
    at transverse coordinate $\bm{x}$ in the direct amplitude and,
    respectively, at $\bm{y}$, in the complex conjugate amplitude.
    This effectively builds up a gluon--gluon dipole of size $\bm{r}
    =\bm{x-y}$, which multiply scatters off the strong color fields
    in the target (represented in the lower part of the figure).
    The evolution of the target wavefunction included Pomeron
    merging and splitting, hence Pomeron loops.
    \label{Ploop} \vspace*{.3cm}}
    \end{figure}

For definiteness, and also for more similarity with the physically
interesting case of $pp$ collisions at LHC, we shall consider that
the collision is viewed in the center--of--mass frame, and that the
two partaking hadrons are identical in that frame --- or, more
precisely, the {\em look} identical when probed via particle
production at {\em central} pseudo--rapidity\footnote{The
pseudo--rapidity is defined as $\eta=-\ln\tan(\theta/2)$, with
$\theta$ the angle between the direction of the produced particle
and the collision axis. Our conventions are such that `forward
rapidities', i.e., positive values for $\eta$, correspond to
particles emitted in the fragmentation region of the projectile. For
a produced gluon, the pseudo--rapidity coincides with the standard
rapidity.}, $\eta=0$. However, differences between the target and
the projectile will appear when one considers particle production at
{\em forward} (pseudo)rapidity $\eta > 0$, since in that case one
probes rather different gluon evolutions in the respective
wavefunctions. Specifically, a gluon emerging with rapidity $\eta$
and transverse momentum $k_\perp$ has been produced via the fusion
of a projectile gluon having longitudinal momentum fraction
$x_1=(k_\perp/\sqrt{s}){\rm e}^{\eta}$ with a target gluon for which
$x_2=(k_\perp/\sqrt{s}){\rm e}^{-\eta}$. We deduce that $y_1\equiv
\ln(1/x_1) =Y/2-\eta$ and similarly $y_2\equiv \ln(1/x_2)
=Y/2+\eta$, with $Y=\ln(s/k_\perp^2)$. Thus, when both $Y/2$ and
$\eta$ are large enough, but not very different from each other, one
can reach the interesting situation where the projectile looks
dilute, while the target is dense.  More precisely, we shall work
under the assumptions that
  \be\label{Ycond}
  Y\,,\ \eta\,\gg \,\frac{1}{\bar\alpha_s}\,\qquad{\rm but}
  \qquad \frac{Y}{2}\,-\,\eta\,\ll \,
 \frac{1}{\bar\alpha_s}\,\ln \frac{1}{\alpha_s^2}\,\,,\ee
($\bar\alpha_s\equiv \alpha_s N_c/\pi$) where the last condition
\cite{AM94,IM03} ensures that saturation effects remain negligible
in the (relevant part of the) projectile wavefunction. In fact, the
most interesting regime for us here is the high--energy regime at
$\bar\alpha_s Y > \ln(1/\alpha_s^2)$ (and similarly for
$\bar\alpha_s\eta$), in which the target wavefunction is
sufficiently evolved to be sensitive to Pomeron loops.

Although there is no real need to further specify our target and
projectile --- since, in the interesting regime at high energy, the
cross--section for gluon production is given by a rather universal
formula (see Sect. \ref{SECT_SIGMA}) --- it is still useful to have
a more specific physical picture at hand, at least at intermediate
stages. To that aim, we shall assume that the incoming hadronic
systems have been produced via the high--energy evolution of an
initial quark--antiquark ($q\bar q$) dipole of size $r_0$ (within
the large--$N_c$ approximation). Then, the wavefunction of the
projectile at rapidity $y_1$ can be described as an {\em onium}
\cite{AM94}, i.e., a collection of $q\bar q$ dipoles which has been
generated in the course of the evolution via $1\to 2$ dipole
splitting. On the other hand, the target wavefunction at rapidity
$y_2$ develops saturation effects, so its description transcends the
dipole picture, even at large $N_c$.

Gluon production in onium--hadron scattering can then be described
as follows (see also Fig. \ref{Ploop}) : a gluon radiated by any of
the dipoles within the onium undergoes multiple scattering off the
strong color fields in the hadron before eventually emerging with
rapidity $\eta$ and transverse momentum $k_\perp\equiv k$. The
differential cross--section for gluon production (or `gluon
spectrum') can be given a factorized structure
\cite{KM98,KTS99,DM01,KT02,KW02,BGV04,CM04,Kov05}:
 \be \f{\rmd\sigma}{\rmd\eta\rmd^2k  \rmd^2b}&=&\f{2\alpha_sC_F}{\pi
 k^2} \int\f{\rmd^2r_1}{2\pi
 r^2_1}\ {n}(r_0,r_1,y_1)\,\nn&{}&\qquad
 \int_0^{r_1} \rmd r \, J_0(kr)\,\log\lr{\f{r_1}r}\,
 \f{\partial}{\partial r}\lr{r \f{\partial}{\partial r}\, \left\bra
 T_{gg}(r)\right\ket_{y_2}}\label{FACT}\ee
which can be heuristically understood with reference to Fig.
\ref{Ploop}. The dimensionless quantity ${n}(r_0,r_1,y_1)$ refers to
the onium wavefunction: it represents the density of dipoles of size
$r_1$ produced from the  original dipole $r_0$ after a BFKL
evolution up to rapidity $y_1=Y/2-\eta$, with the initial condition
${n}(r_0,r_1,0)=r_0 \delta({r_1}-{r_0})$. Furthermore,
$\alpha_sC_F\ln(r_1/r)$ can be roughly interpreted as the
probability that an elementary dipole of size $r_1$ radiate a
gluon--gluon ($gg$) dipole of size $r$, the dipole being effectively
constructed with a gluon line in the direct amplitude times another
one in the complex conjugate amplitude ($r$ is the transverse
distance between these two lines).  $\bra T_{gg}(r)\ket_{y_2}$ is
the (average) scattering amplitude for the collision between the
$gg$ dipole and the hadronic target evolved up to rapidity
$y_2=Y/2+\eta$. Finally, the Bessel function $J_0(kr)$ together with
the integral over $r$ implements the Fourier transform from the
dipole size $r$ to the transverse momentum $k$ of the produced
gluon. The presence of the derivatives w.r.t. $r$ in
Eq.~(\ref{FACT}), for which we have not given any interpretation,
makes it clear that our above `physical interpretation' is rather
crude. A better understanding of Eq.~(\ref{FACT}) would require a
careful inspection of its derivation in the literature, but this
goes beyond our purposes here.

Eq.~(\ref{FACT}) is often referred to as expressing
``$k_\perp$--factorization'', and it is instructive to understand
why. To that aim, it is convenient to first solve the BFKL equation
for ${n}(r_0,r_1,y_1)$ and then insert the result into
Eq.~(\ref{FACT}). One thus finds
 \be
 \f{\rmd\sigma}{\rmd\eta\rmd^2k  \rmd^2b}
 =\f{\abar}{ k^2} \int_0^\infty
 \rmd r\, J_0(kr)\,\tilde{n}(r_0,r,y_1)\, \f{\partial}{\partial r}
 \lr{r\f{\partial}{\partial r} \left\bra
 T_{gg}(r)\right\ket_{y_2}}\label{bfkl}\ee
(we have used $2C_F\approx N_c$ at large $N_c$ together with
$\abar\equiv \alpha_s N_c/\pi$), where
 \be\label{tilden}
\tilde{n}(r_0,r,y)\equiv \int_C\frac{\rmd\gamma}{2\pi i}
\,\lr{\f{r_0}r}^{2\gamma}\f1{2\gamma^2}
 \exp\big({\abar}y\chi(\gamma)\big)\ ,\ee
is essentially the Mellin representation of the BFKL solution.
($\chi(\gamma)$ is the standard BFKL characteristic function
\cite{BFKL} and the contour $C$ in the complex $\gamma$ plane is
chosen as $C =\left\{\gamma=\gamma_1 + i \gamma_2\,;-\infty<
\gamma_2 <\infty\right\}$, with a fixed $\gamma_1$ in the range $0 <
\gamma_1 < 1$.) Eq.~(\ref{bfkl}) is finally rewritten in a form on
which $k_\perp$--factorization becomes manifest :
 \be\label{kTgluon}
 \f{\rmd\sigma}{\rmd\eta\rmd^2k  \rmd^2b}
 =\f{\abar}{k^2} \int\frac{\rmd^2 {\bm p}}{(2\pi)^2}\,
 \varphi({\bm p},y_1) \Phi(\bm{k-p},y_2)\,,\ee
where
 \be\label{phip}
 \varphi({\bm p},y)\equiv  \int\f{\rmd^2\bm{r}}{2\pi}
 \ {\rm e}^{i\bm{p\cdot r}}\,\tilde{n}(r_0,r,y)\,\ee
represents (up to a normalization factor of order $\alpha_s$) the
unintegrated gluon distribution in the onium, whereas the quantity
 \be\label{Phidef}
 \Phi(\bm{k},y)\equiv  \int{\rmd^2\bm{r}}\ {\rm e}^{i\bm{k\cdot r}}\,
 \grad^2_{\bm{r}} \left\bra T_{gg}(\bm{r})\right\ket_{y} \ee
can be similarly interpreted as a {\em generalized\,} `unintegrated
gluon distribution' for the target, which is however sensitive to
the non--linear effects in the target (related to gluon saturation)
via the multiple scattering encoded in the dipole amplitude. It is
only for sufficiently large $k$ --- much larger than the target
saturation momentum --- that the quantity
$(1/\alpha_s)\Phi(\bm{k},y)$ reduces (up to a numerical factor) to
the standard, leading--twist, unintegrated gluon distribution. Note
also that the two distributions defined in Eqs.~(\ref{phip}) and
(\ref{Phidef}) have different dimensions: this is so since, whereas
the onium distribution $\varphi({\bm p},y)$  has been already
integrated over the transverse area of the onium (so this is
`unintegrated' only with respect to ${\bm p}$), the target
distribution $\Phi(\bm{k},y)$ on the other hand is rather understood
at a fixed impact parameter ${\bm b}$ within the area covered by the
target (hence, this is more properly a {\em gluon occupation factor}
in the transverse phase--space).

In previous applications of Eqs.~(\ref{FACT})--(\ref{Phidef}), one
has always treated the wavefunction of the dense hadronic target in
the {\em mean field approximation}, that is, one has included the
effects of saturation and multiple scattering --- e.g., by computing
the dipole scattering amplitude $\bra T_{gg}(\bm{r})\ket_{y}$ with
the Glauber--Mueller formula, or by solving the BK equation --- but
one has ignored the gluon--number fluctuations (although the
applicability of the factorization (\ref{FACT}) in the presence of
Pomeron loops has been noticed\footnote{Note, however, that the
analysis in Ref. \cite{Kov05} includes only the `large' Pomeron
loops which appear when the collision between the $gg$ dipole and
the target is viewed as onium--onium scattering in the
center--of--mass (COM) frame --- here, the COM of the subsystem made
by the $gg$ dipole and the target. Thus, this analysis is limited to
not too high energies \cite{AM94,IM03}, unlike our present approach
where the Pomeron loops are included directly in the target
wavefunction.} by Kovchegov \cite{Kov05}). Our purpose in what
follows is to go beyond such previous analyses, by including the
effects of the Pomeron loops in the calculation of $\bra
T_{gg}(\bm{r})\ket_{y}$. As we shall see, these effects lead to a
dramatically new physical picture at sufficiently high energy.

The Pomeron loop equations are written for dipoles made with
quark--antiquark pairs (rather than gluons), but this is not a
problem since at high energy the scattering amplitudes for the two
types of dipoles ($gg$ and $q\bar q$) are simply related to each
other. Namely, if $S_{gg}$ and $S_{q\bar q}$ denote the respective
$S$--matrices, with $T=1-S$, then in the eikonal approximation we
have:
 \be S_{gg}(r) \,=\,\frac{N_c^2}{N_c^2-1}\,S_{q{\bar q}}^2(r)\,\approx\,
 S_{q\bar q}^2(r)\,,\ee
where the last, approximate, equality holds at large $N_c$, and we
have also assumed that the $S$--matrix is real, as appropriate at
high energy; this in turn implies:
 \be\label{gq}
  T_{gg}(r)\,\equiv\,1- S_{gg}(r) \,\approx\,
 2T_{q\bar q}(r)-T_{q\bar q}^2(r)\,.\ee
Notice that, in the previous formul\ae, we have never indicated the
dependence of the dipole amplitude upon the dipole impact parameter,
but only upon its size. This is for consistency with our subsequent
approximations, in which we shall systematically neglect the
$b$--dependence, i.e., we shall compute amplitudes or
cross--sections at fixed impact parameter and assume the
high--energy evolution to be quasi--local in $b$.

In the next section we shall concentrate ourselves on the $q\bar q$
amplitude $T_{q\bar q}(r)$ --- that we shall denote simply as $T(r)$
from now on --- and recall some recent results concerning the
calculation of this amplitude from the Pomeron loop equations. The
consequences of these results for the generalized gluon distribution
(\ref{Phidef}) and for the cross--section (\ref{FACT}) for gluon
production will be then examined in the subsequent sections.

\section{The dipole amplitude: Black spots \& Diffusive scaling}
\setcounter{equation}{0}\label{SECT_DIPOLE}

If there was not for (gluon--number) fluctuations, i.e., in the mean
field approximation to the non--linear evolution in QCD at high
energy, the dipole amplitude $T(r,Y)$ at large $N_c$ would be given
by the solution to the BK equation \cite{B,K} --- a non--linear
generalization of the BFKL equation which is consistent with
unitarity. Even in the presence of fluctuations, the BK amplitude
remains a reasonable approximation for the {\em event--by--event}
amplitude, and also for the {\em average} amplitude $\bra
T(r)\ket_Y$ at {\em intermediate} values of $Y$. It is therefore
useful to begin our discussion with a brief reminder of the BK
solution.

{\bf I)} Although the latter is not known {\em exactly}, its basic
properties are well understood --- thanks to a multitude of analytic
and numerical studies
\cite{Motyka,LT99,LL01,SCALING,MT02,MP03,RW03,Nestor03,MS05} --- and
a piecewise analytic approximation to it can be written down. To
that purpose, it is preferable to measure the dipole size in
logarithmic units, by writing:
 \be\label{Tev1}
 T(r,Y) \,\equiv  \, T(\rho,Y)\,\qquad{\rm with}\qquad
 \rho\equiv \ln \frac{r_0^2}{r^2}\,,\ee
where $r_0$ is the unitarization scale in the target at low energy
(e.g., if the target starts as a bare dipole at $Y=0$, then $r_0$
is the size of that dipole). Note that large values of $\rho$
correspond to small dipole sizes, or to large transverse momenta
($\rho\sim \ln k^2$) after a Fourier transform.

When viewed as a function of $\rho$ for a fixed (and sufficiently
large) $Y$, the BK solution $T(\rho,Y)$ appears as a {\em front}
which interpolates between the unitarity (or `black disk') limit,
$T=1$, at relatively small values of $\rho$ and `color
transparency', $T\sim {\rm e}^{-\rho}$, at very large values of
$\rho$, with the transition between the two regimes (`the front
region') governed by the BFKL dynamics with saturation boundary
conditions (see Eq.~(\ref{SBC}) below). The {\em position} of the
front, conventionally defined as the value $\rho_s(Y)$ at which
$T=1/2$, represents the unitarization scale at rapidity $Y$, i.e.,
the {\em saturation momentum} $Q_s(Y)$ \cite{GLR,MV} in logarithmic
units. Studies of the BK equation \cite{MP03} or, simpler, of the
BFKL equation supplemented with a saturation boundary condition
\cite{SCALING,MT02}, which reads :
 \be\label{SBC}
 T(\rho,Y)\,=\,1/2 \,\qquad{\rm for}\qquad
 \rho = \rho_s(Y)\equiv \ln \big(r_0^2Q_s^2(Y)\big)\,,\ee
reveal that, for large $Y$, the front propagates at constant
speed: $\rho_s(Y) = \lambda_0 \bar\alpha_s Y$, with $\lambda_0
\approx 4.88$. Equivalently, the saturation momentum $Q_s(Y)$
rises exponentially with $Y$ \cite{GLR}, with a `saturation
exponent' $\lambda_0 \bar\alpha_s$.

The most interesting region for us here is the front region at
(roughly) $\rho_s(Y)< \rho < 2\rho_s(Y)$, where the scattering is
weak ($T \ll 1$), yet it is significantly influenced by
saturation, via the boundary condition (\ref{SBC}). Within this
region, the BK amplitude shows approximate {\em geometric
scaling}, i.e., it depends upon the two kinematical variables
$\rho$ and $Y$ only via the difference $z\equiv \rho-\rho_s(Y)
=\ln[1/r^2Q_s^2(Y)]$. More precisely, one finds
\cite{SCALING,MT02}
 \be\label{Texp}
 T(z,Y) \, \sim \,z\,
 {\rm e}^{-\gamma_0 z}\,
 \exp\left\{-\frac{z^2}
 {2\beta_0 \bar\alpha_s Y} \right\}\qquad{\rm for}\qquad
 1 \ll z\equiv \rho-\rho_s(Y) \ll \rho_s(Y).
 \ee
This expression involves the `anomalous dimension'
$\gamma_0\approx 0.63$ and the BFKL `diffusion coefficient'
$\beta_0\equiv \chi''(\gamma_0)\approx 48.52$, which are both
hallmarks of the BFKL dynamics in the presence of saturation. Note
that the last factor in Eq.~(\ref{Texp}), describing {\em BFKL
diffusion}, depends separately upon $z$ and $Y$, hence it violates
geometric scaling. Still, when increasing $Y$, there is an
increasing domain in $z$
--- the {\em geometric scaling window} at $1\simle z \simle
z_g(Y)$, with $z_g(Y) \sim (2\beta_0 \bar\alpha_s Y)^{1/2}$ ---
within which BFKL diffusion can be neglected, and the BK amplitude
shows good geometric scaling: $T\sim z\, {\rm e}^{-\gamma_0 z}$.
On the other hand, for much larger value of $z\simge \rho_s(Y)$,
one enters the regime of color transparency, where the BK solution
reduces to the standard `double--logarithmic approximation'. This
is the cross--over domain towards the DGLAP dynamics and collinear
factorization.

One should stress at this point that all these remarkable
predictions of the BK equation --- the exponential growth of the
saturation momentum with $Y$, the existence of a window for
geometric scaling at moderately high $Q^2\gg Q_s^2(Y)$, and the
violation of this scaling at even higher values of $Q^2$ ---
appear to be necessary in order to explain the HERA data at
small--$x$, and the RHIC data for particle production in d+Au
collisions at forward rapidities (cf. the discussion in the
Introduction). Remarkably, these properties of the mean field
approximation are qualitatively preserved after including the NLO
corrections to the BFKL equation, but they suffer quantitative
modifications \cite{DT02}, which have the right trend to achieve a
better description of the data \cite{IIM03}.

{\bf II)} The basic properties of the BK solution are also preserved
by fluctuations, but only at the level of the {\em event--by--event}
amplitude, by which we mean the amplitude obtained after a single
realization of the stochastic evolution. Namely, this amplitude can
be given the following piecewise approximation \cite{IMM04}
 \begin{equation}\label{Tevent}
    T(z)=
    \begin{cases}
        \displaystyle{1} &
        \text{ for\,  $z < 0$}
        \\*[0.2cm]
        \displaystyle{
        {\rm e}^{-\gamma_0 z}} &
        \text{ for\,  $1 < z < L$}
        \\*[0.2cm]
        \displaystyle{
        {\rm e}^{- z}} &
        \text{ for\,  $z \gg L $}.
    \end{cases}
 \end{equation}
where $z\equiv \rho - \rho_s(Y)$ denotes the geometric scaling
variable, as before, and we have kept only the dominant,
exponential, behaviour in each region. As compared to the BK
amplitude discussed previously, there are two important
differences though:

\texttt{a)} The front velocity $\lambda$, which implicitly enters
Eq.~(\ref{Tevent}) via $\rho_s\equiv \lambda \bar\alpha_s Y$, is
considerably smaller than the corresponding prediction of the BK
equation, $\lambda_0\approx 4.88$. The difference
$\lambda_0-\lambda$ is analytically under control only in the
(physically unrealistic) weak coupling limit %$\alpha_s^2\to 0$
\cite{BD97,MS04,IMM04}. On then obtains
 \begin{equation}
 \lambda\,\simeq\,\lambda_0\,-\, \frac{\cal
 C}{\ln^2(1/\alpha_s^2)}\qquad {\rm when}\qquad \alpha_s^2\to 0\,,
 \label{satscal}
 \end{equation}
where the coefficient ${\cal C}$ turns out to be quite huge: ${\cal
C}=\pi^2 \gamma_0 \chi''(\gamma_0)\approx 150$. Note that, when
$\alpha_s^2\to 0$, $\lambda$ converges towards $\lambda_0$ (from the
below), but {\em only very slowly} (logarithmically).
%Because of the theoretical uncertainty in the value of $\lambda$ for
%realistic values of $\alpha_s$, we shall
The tendency of $\lambda$ to decrease when increasing $\alpha_s^2$
is also confirmed by a study of the `strong noise limit', which
shows that in that limit $\lambda$ vanishes as a power of
$1/\alpha_s^2$ \cite{MPS05}.

\texttt{b)} The front region, within which the amplitude
(\ref{Tevent}) exhibits geometric scaling with the BFKL `anomalous
dimension' $\gamma_0\approx 0.63$, is {\em compact}, i.e., it has a
finite width $L$ which is independent of $Y$. Specifically, $L$ is
the distance $\rho - \rho_s$ over which the amplitude falls off from
its saturation value $T=1$ to a value of order $\alpha_s^2$, where
the fluctuations wash out completely any mean field behaviour. This
condition immediately yields:
 \be\label{L}
  L\,\simeq\,\frac{1}{\gamma_0}\,\ln\,\frac{1}{\alpha_s^2}
  \,+ \,O(1)\,.
 \ee
More precisely, in the early stages of the evolution, the width of
the geometric scaling window increases via BFKL diffusion (cf.
Eq.~(\ref{Texp})), until it gets stuck at its maximal value, equal
to $L$. The condition $z_g(Y_{\rm form})=L$ determines the front
formation rapidity as
   \be\label{Dtau}
 \bar\alpha_s \, Y_{\rm form}
 \,\sim\,
 \frac{\ln^2 (1/\alpha_s^2)}{2 \beta_0\,\gamma_0^2}\,.\ee
%For larger rapidities $Y > Y_{\rm form}$, the BFKL diffusion plays
%no role anymore.

{\bf III)} But the most spectacular effects of fluctuations appear
in the calculation of {\em expectation values}, so like the {\em
average} amplitude $\lan T(\rho) \ran_Y$ : for sufficiently large
values of $Y$, this averaged quantity loses any memory of the shape
of the event--by--event front (in particular, of geometric scaling),
since it is dominated by {\em front dispersion} \cite{IT04}.

Specifically, starting with an unique initial condition at $Y=0$, a
stochastic evolution generates an {\em ensemble of fronts} at
rapidity $Y$, where the individual fronts have more or less the same
shape (cf. Eq.~(\ref{Tevent})) but differ from each other by a
translation. That is, the ensemble can be described by treating the
position $\rho_s$ of a front as a {\em random variable}, which on
the basis of the correspondence with statistical physics can be
argued to have a Gaussian probability distribution
\cite{BD97,IMM04,BDMM} :
 \begin{equation}\label{probdens}
    P_Y(\rho_s) =
    \frac{1}{\sqrt{\pi}\sigma}\,
    \exp \left[
    -\frac{\left( \rho_s - \langle \rho_s \rangle \right)^2}{\sigma^2}
    \right].
 \end{equation}
with
 \be
 \langle \rho_s \rangle\,=\,\lambda\bar\alpha_s Y,\qquad {\rm and}
 \qquad \sigma^2\equiv 2\big(\langle\rho_s^2\rangle -
 \langle\rho_s\rangle^2\big) \,=\,D_{\rm fr}\bar\alpha_s Y,\ee
where the {\em average front velocity} $\lambda$ and the {\em
front diffusion coefficient} $D_{\rm fr}$ are analytically known
only in the limit $\alpha_s^2\to 0$: $\lambda$ is then given by
Eq.~(\ref{satscal}), whereas $D_{\rm fr}$ scales as
 \be\label{Dfr} D_{\rm fr}\,\simeq\,\frac{\cal
 D}{\ln^3(1/\alpha_s^2)} \qquad {\rm when}\qquad \alpha_s^2\,\to\, 0\,,\ee
with a coefficient ${\cal D}$ that has been recently computed in
Ref. \cite{BDMM}. Since the parameters $\lambda$ and $D_{\rm fr}$
not known for realistic values of $\alpha_s$, in what follows we
shall treat them as {\em free parameters}. In fact, rather than
$\rho$ and $Y$, it will be more convenient to work with the scaled
variables $z\equiv \rho - \langle \rho_s \rangle$ and $\sigma$.

\begin{figure}[t]\hspace*{-1.cm}
    \centerline{\epsfxsize=10.cm\epsfbox{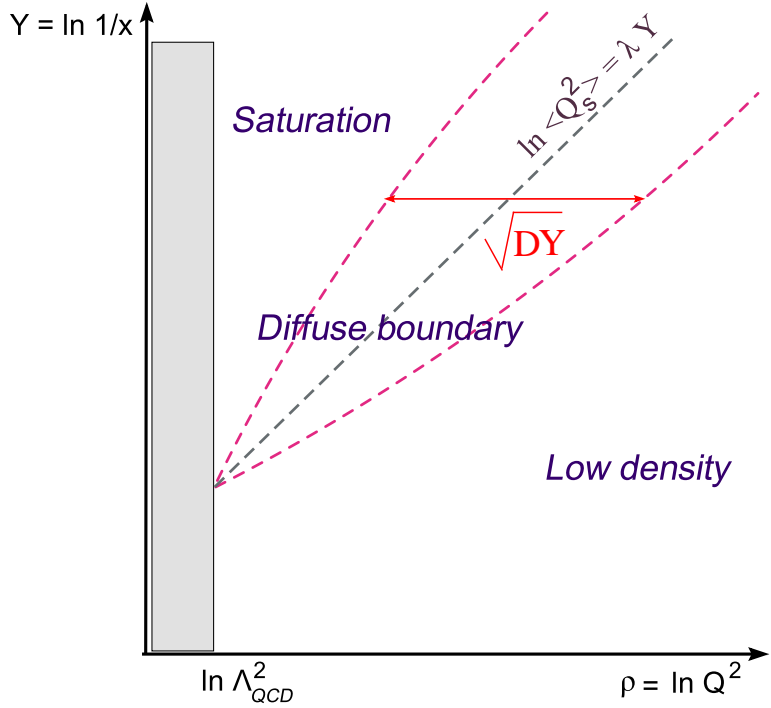}}
    \caption{\sl  The diffusive saturation boundary generated by
    the evolution with fluctuations.
    \label{phaseDiff}}\bigskip
    \end{figure}

We see that, as a consequence of fluctuations, the saturation
boundary become {\em diffuse}, within a radius $\sigma\propto
\sqrt{Y}$ around the {\em average} saturation line $\langle \rho_s
\rangle$ (see Fig. \ref{phaseDiff}). This mechanism should not be
confused with the BFKL diffusion, which applies to the individual
fronts, and which becomes anyway inoperative for sufficiently large
rapidities $Y > Y_{\rm form}$, as we have seen. The dispersion in
the saturation boundary rather means that, in the vicinity of the
average saturation line $\langle \rho_s \rangle$, the hadron
wavefunction is the site of wild fluctuations in the gluon density.
When probing the hadron (say, with an external dipole) on a
resolution scale $\rho$ within the diffusive radius ($|\rho -
\langle \rho_s \rangle|\simle \sigma$), one can either find a `black
spot' --- a gluon configuration with saturation scale $\rho_s \ge
\rho$, which will completely absorb the projectile ---, or a `white
spot' --- a configuration having $\rho_s \ll \rho$, for which the
dipole will emerge unscattered
---, or any intermediate nuance of `grey'.

\begin{figure}[t]%\hspace*{-1.cm}
    \centerline{\epsfxsize=15.cm\epsfbox{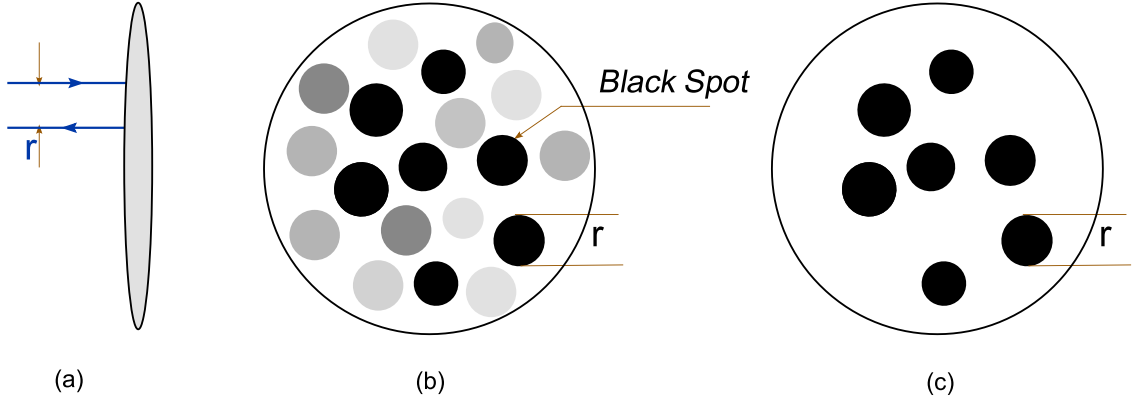}}
    \caption{\sl Dipole--hadron scattering in the
    fluctuation--dominated regime at $\sigma^2\gg 1$.
    (a) a view along the collision axis; (b)
    a transverse view of the hadron, as `seen' by a small dipole
    impinging at different impact parameters; (c) the simplified,
     black\&{white}, picture of the hadron which is relevant for
     the average dipole amplitude.
    \label{HotSpot}}\bigskip
    \end{figure}

Although this picture has been obtained by working at a fixed impact
parameter, it can be easily extended to a picture of the target in
impact parameter space, under the assumption that the high--energy
evolution is {\em quasi--local} in $b$ (as argued in Refs.
\cite{IMM04,IT04}). Then, the evolutions at different points $b$
proceed (quasi)independently from each other and randomly generate
any of the fronts which compose the statistical ensemble at $Y$.
Accordingly, the average over the ensemble of fronts is tantamount
to an average over all the impact parameters within the target area.
We see that, even if one starts with homogeneous initial conditions
at $Y=0$, the target becomes highly inhomogeneous at large $Y$, as a
consequence of (gluon--number) fluctuations in the course of the
evolution. These inhomogeneities can be probed by a dipole
projectile: A dipole with size $r$ explores an area $\sim r^2$
around its impact parameter $b$. Because of the strong dispersion in
the ensemble of fronts at large $Y$, a small dipole with $\rho \gg
\langle \rho_s \rangle$ can `see' either `black spots', or `white',
or `grey', depending upon its impact parameter. This is illustrated
in Figs. \ref{HotSpot}.a and b.

The crucial feature about this situation is the fact that, for
sufficiently large values of $Y$ (such that $\sigma\gg 1$), the
`black spots' completely dominate the average dipole amplitudes
\cite{IT04} up to very large values of $\rho \gg \langle \rho_s
\rangle$, that is, including in the `weak scattering regime' where
$\lan T(\rho) \ran_Y\ll 1$. When this happens, the contribution of
the `grey spots' is negligible, so the hadron appears to the
external dipole as {\em either black, or white} (see Fig.
\ref{HotSpot}.c) : mostly black when $\rho \ll \langle \rho_s
\rangle$, but mostly white in the opposite situation where $\rho \gg
\langle \rho_s \rangle$. In fact, for $\sigma\gg 1$ and at all the
points deeply inside the diffusion radius ($|\rho - \langle \rho_s
\rangle|\ll \sigma$), the proportions of black and white are almost
the same, so the average scattering amplitude is simply one--half !
But the dominance of the black spots actually extends up to much
larger values of $\rho$, well outside the diffusive radius, namely
so long as $\rho - \langle \rho_s \rangle \ll \sigma^2$.

To understand how this radically new physical picture arises,
consider the average dipole amplitude, which in the present context
is obtained as an average over the statistical ensemble of fronts,
with weight function (\ref{probdens}):
 \begin{equation}\label{Tavedef}
    \lan T(\rho) \ran_Y =
    \int \limits_{-\infty}^{\infty}
    \rmd\rho_s\, P_Y(\rho_s)\, T(\rho -\rho_s),
\end{equation}
with $T(\rho -\rho_s)$ the single--event front in
Eq.~(\ref{Tevent}).  (Higher--point correlations can be similarly
computed.) The values of $\rho_s$ which dominate this convolution
depend upon the competition between $\sigma$ (the width of the
Gaussian distribution of the fronts) and $1/\gamma$, with $\gamma
\sim {\cal O}(1)$,  which characterizes the exponential decay of
the individual fronts. Specifically:

\texttt{i)} When $\sigma \ll 1/\gamma$ (the situation in the early
stages of the evolution), the Gaussian ensemble is strongly peaked
around the average front, hence the average amplitude retains the
single event profile:
 \be \label{Tsingle}
    \lan T(\rho) \ran_Y \,\approx\,T(\rho -\langle
  \rho_s \rangle)\,,\ee
and thus shows {\em geometric scaling}.

\texttt{ii)} When $\sigma \gg 1/\gamma$ (the situation at
sufficiently high energy), then for all the values of $\rho$ such
that $z\equiv \rho - \lan \rho_s \ran \ll \gamma\sigma^2$, the
average amplitude is dominated by `black spots', i.e., by
configurations with $\rho_s\ge \rho$, which look black ($T=1$) on
the resolution scale of the projectile. Note that, for large $\rho
> \lan \rho_s \ran$, such configurations are relatively rare, yet
they dominate the convolution (\ref{Tavedef}) since the
contributions  $T(\rho -\rho_s)$ of the typical configurations
(for which $\rho_s\sim  \lan \rho_s \ran$) are exponentially
suppressed. One then finds
 \be\label{Thighsigma}
    \lan T (\rho)\ran_Y \,\simeq\,
    \int \limits_{\rho}^{\infty}
    \rmd\rho_s\, P_Y(\rho_s)\,\,=\,
    \frac{1}{2}\, {\rm Erfc}\left(\frac{z}{\sigma} \right)
    \qquad {\rm for} \quad -\infty < z \ll \gamma\sigma^2,
 \ee
where the neglected terms are suppressed by, at least, one power
of $1/\sigma$ and/or $z/\sigma^2$. (Note that we often use the
fact that $\gamma\sim O(1)$ to simplify the parametric estimates.)
Eq.~(\ref{Thighsigma}) involves the complementary error function,
 \be \label{erfcdef}
    {\rm Erfc}(x)\,\equiv\,\frac{2}{\sqrt{\pi}}
    \int\limits_x^\infty {\rm
    d}t\,{\rm e}^{-t^2} \,=\
    \begin{cases}
        \displaystyle{2-\frac{\exp(-x^2)}{\sqrt{\pi}\,|x|}} &
        \text{ for\,  $x \ll -1$}
        \\*[0.1cm]
        \displaystyle{1} &
        \text{ for\,  $|x|\ll 1$}
        \\*[0.1cm]
        \displaystyle{\frac{\exp(-x^2)}{\sqrt{\pi}x}} &
        \text{ for\,  $x \gg 1$}\,.
    \end{cases}
 \ee
As anticipated, the average amplitude at high energy  is
approximately constant, $\lan T (\rho)\ran_Y  \approx 1/2$, within
the large interval $|z|\ll \sigma$ around the average saturation
line. Moreover, within the whole validity region of the
approximation (\ref{Thighsigma}), i.e., for $z\ll\sigma^2$, the
amplitude shows {\em diffusive scaling} : it depends upon $\rho$ (or
$r$) and $Y$ only via the single variable
  \be\label{taudef}
  \tau\,\equiv\,\frac{z}{\sigma}\,\equiv\,
  \frac{\rho - \lan \rho_s \ran}{\sigma}\,\equiv\,\frac{\ln(1/r^2
  \langle Q_s^2 \rangle)}{\sigma}\,.\ee
The physical regions for high--energy evolution in the kinematical
plane $\rho-Y$ are illustrated in Fig. \ref{phase}. The boundary of
the geometric scaling window at large $\rho$ follows from the
discussion after Eq.~(\ref{Tevent}). As for the diffusive scaling
window, this extends up to $\rho_{\rm max}\simeq \langle \rho_s
\rangle + \sigma^2 =(\lambda+D_{\rm fr})\bar\alpha_s Y$, cf.
Eq.~(\ref{Thighsigma}). A more complete expression for the average
amplitude which interpolates between these various regions will be
presented in the next section (see Eq.~(\ref{Tave})). Here, we would
like to focus on the high--energy amplitude (\ref{Thighsigma}),
which exhibits some other remarkable properties:

\begin{figure}[t]\hspace*{-0.5cm}
    \centerline{\epsfxsize=13.2cm\epsfbox{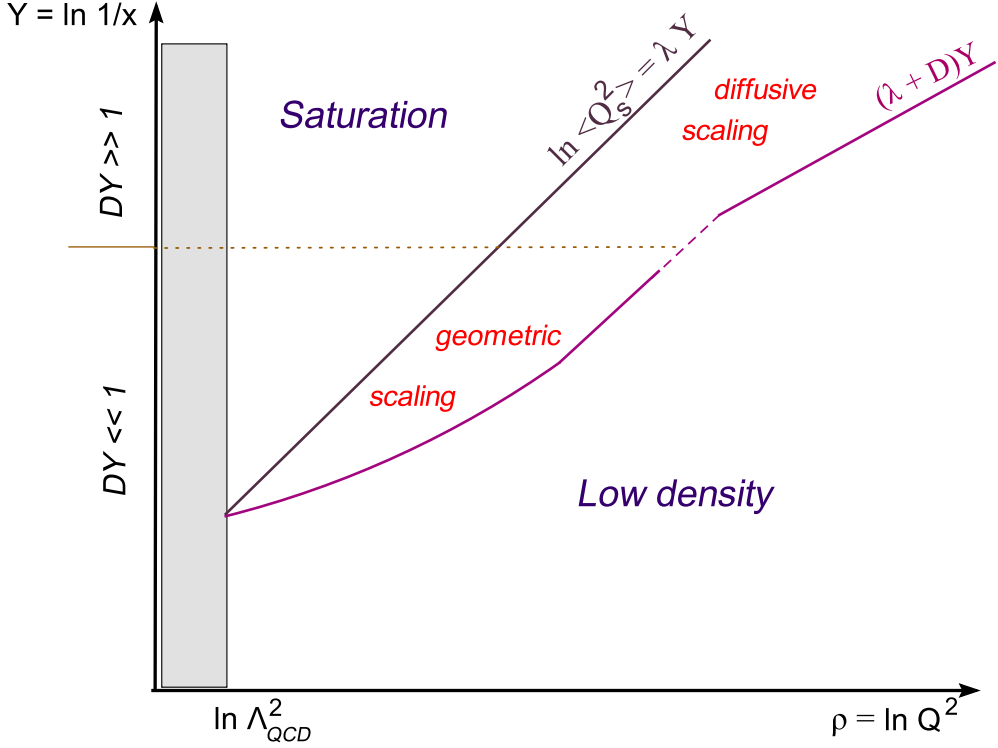}}
    \caption{\sl A `phase--diagram' for the high--energy evolution
    with Pomeron loops. Shown are the average saturation line and
    the approximate boundaries of the scaling regions at large
    values of $\rho\sim \ln Q^2$ (with the simplified notations
    $\langle \rho_s \rangle =\lambda Y$ and $\sigma^2 = DY$).
    Note the gradual transition with increasing $Y$ from
    geometric scaling at intermediate energies ($DY\ll 1$)
    to diffusive scaling at very high  energies ($DY\gg 1$).
    This transition can be more quantitatively studied on
    Eq.~(\ref{Tave}).
    \label{phase}}
    \end{figure}

For relatively large $\rho\gg\lan \rho_s \ran$ such that $z\gg
\sigma$ (with $z\ll \sigma^2$ though), this function decreases as a
Gaussian in $z$, rather than as an exponential. Thus, within a
rather wide window at high $Q^2$, whose width is increasing with $Y$
(see Fig. \ref{phase}), the scattering is weak, yet there is {\em no
`twist--expansion'}
--- i.e., no expansion in powers of $r^2\sim 1/Q^2$ --- for the
scattering amplitude. `Color transparency', meaning that $\lan T
(\rho)\ran \propto {\rm e}^{-\rho}$, is eventually recovered, but
only for very large $z \simge \sigma^2$  (cf. Eq.~(\ref{Tave})).
Furthermore, the amplitude (\ref{Thighsigma}) shows {\em no
`Pomeron'--like growth} --- i.e., no power--like increase with $Y$
at fixed $\rho$.

Such properties might look curious,  but in fact they are easy to
understand in the present context: Both color transparency and the
power--like increase with $Y$ are properties of the scattering
between the projectile dipole and a {\em dilute} gluon configuration
in the target. But as already explained, Eq.~(\ref{Thighsigma})
rather describes the scattering off {\em dense} gluon
configurations, for which the individual amplitudes have already
reached the unitarity limit $T=1$. When varying $\rho$ or $Y$, we
modify the composition of the statistical ensemble which is probed
by the scattering, but the average amplitude changes only slowly.

It is important to emphasize that the functional form
(\ref{Thighsigma}) of the high--energy amplitude  is {\em extremely
robust}, as it follows from very few and general physical
assumptions: the facts that the amplitude $T(r)$ saturates the
unitarity limit $T=1$ for sufficiently large dipole sizes $r$ and
vanishes like a power of $r$ when $r\to 0$, and that the
gluon--number fluctuations in the target lead to a dispersion in the
ensemble of fronts, which increases with $Y$. By the same arguments,
this result is {\em universal} : it is insensitive to the initial
conditions at low energy and also to the details of the high--energy
evolution. The latter matter only for the calculation of the
parameters $\lambda$ and $D_{\rm fr}$, or, more generally, of the
$Y$--dependencies of the average saturation scale $\lan \rho_s(Y)
\ran$ and of the front dispersion $\sigma(Y)$.

By using Eq.~(\ref{Thighsigma}) within the dipole factorization
scheme for DIS, we have been able to estimate (in a previous work
which also included Y. Hatta and D. N. Triantafyllopoulos
\cite{HIMST06}) the high--energy limit of the inclusive and
diffractive DIS cross--sections, and thus demonstrate that all the
remarkable properties discussed above in relation with the dipole
amplitude do also transmit to these cross--sections. Returning to
the problem of interest for us here --- namely, gluon production in
onium--hadron collisions at forward rapidity --- it seems natural to
anticipate that these properties will show up in this context too,
since the corresponding cross--section (\ref{FACT}) is again related
to the dipole amplitude. But a closer inspection of Eq.~(\ref{FACT})
may shed serious doubts on this conclusion: the `generalized gluon
occupation factor' defined in Eq.~(\ref{Phidef}), and which is a
building block of the cross--section (\ref{FACT}), involves the
(second order) {\em derivative} of the dipole amplitude. When taking
this derivative on the convolution (\ref{Tavedef}), we kill the
contribution of the fronts which are at saturation in the
statistical ensemble (the `black spots'), because for them $T=1$.
Rather, the whole non--trivial contribution to this derivative
arises from the  {\em tails} of the individual fronts. Thus,
clearly, gluon production is dominated by a different physics as
compared to DIS, and one may wonder whether this physics is
universal at high energy, and whether it can be computed within the
present formalism. The answer to both questions turns out to be
`yes', as we shall argue in the next section.

\section{The generalized gluon distribution}
\setcounter{equation}{0}\label{SECT_GLUON}

We start with Eq.~(\ref{Phidef}) which is rewritten as (after
performing the angular integration)
 \be\label{Phirho}
 \Phi(k,Y)&\,= &\int_0^\infty \rmd r\,r\, J_0(kr)\,
 \frac{1}{r}\,\f{\partial}{\partial r}\lr{r \f{\partial}{\partial r}\,
 \left\bra T_{gg}(r)\right\ket_{Y}}\,\nn &\,=\,&2
 \int\rmd \rho\, J_0\lr{kr_0{\rm e}^{-\rho/2}}\,
 \f{\partial^2}{\partial \rho^2}\lr{2\lan T (\rho)\ran_Y
 - \lan T^2 (\rho)\ran_Y},\ee
where $\rho\equiv \ln ({r_0^2}/{r^2})$ and we have used
Eq.~(\ref{gq}) to express the second line in term of $q\bar q$
dipole amplitudes alone. As explained at the end of the previous
section, the derivative $\partial_\rho \lan T (\rho)\ran_Y$ is
controlled by the tails of the individual fronts in the statistical
ensemble. Hence, in order to compute this quantity, one needs to
evaluate Eq.~(\ref{Tavedef}) with a complete profile for the
single--event front, which includes a model for the tail. However,
we shall shortly argue that the precise shape of the tail is
unimportant in the kinematic range of interest and, moreover, the
required derivatives can be simply obtained by differentiating
Eq.~(\ref{Thighsigma}) !  This seems paradoxical, since
Eq.~(\ref{Thighsigma}) has been {\em a priori} obtained by
neglecting the tails altogether ! This paradox is resolved by
observing that the dominant contributions to these derivatives are
actually given by the {\em end points} of the tails towards the
saturation region (that is, by $\rho\sim\rho_s$).

In order to show this and also compute $\partial_\rho \lan T
(\rho)\ran_Y$ to the desired accuracy, we shall use the following,
simplified, profile for a single front (compare to
Eq.~(\ref{Tevent}))
 \begin{equation}\label{Tstep}
    T(\rho)=
    \begin{cases}
        \displaystyle{1} &
        \text{ for\,  $\rho < \rho_s$}
        \\*[0.2cm]
        \displaystyle{
        {\rm e}^{-\gamma(\rho - \rho_s)}} &
        \text{ for\,  $\rho \ge \rho_s$}\,,
    \end{cases}
 \end{equation}
with generic $\gamma\sim {\cal O}(1)$. A simple calculation yields
(with $z\equiv \rho - \langle \rho_s \rangle$)
 \begin{equation}\label{Tave}
    \lan T (\rho)\ran_Y=
    \frac{1}{2}\, {\rm Erfc}\left(\frac{z}{\sigma} \right)
    +\frac{1}{2}
    \exp\left( \frac{\gamma^2 \sigma^2}{4} - \gamma z\right)
    \left[ 2 - {\rm Erfc}\left(\frac{z}{\sigma} -
    \frac{\gamma \sigma}{2} \right)\right].
\end{equation}
The first term is the same as in Eq.~(\ref{Thighsigma}) and comes
from the saturation piece of the single front (\ref{Tstep}), while
the second term comes from the exponential tail in
Eq.~(\ref{Tstep}). Incidentally, the previous results in
Eqs.~(\ref{Tsingle}) and (\ref{Thighsigma}) can be easily recovered
by expanding Eq.~(\ref{Tave}) in the corresponding limits. The
corresponding expression for $\lan T^2 (\rho)\ran_Y$ is obtained by
replacing $\gamma\to 2\gamma$ into Eq.~(\ref{Tave}).

At this level, it is straightforward to compute the required
derivatives by acting with $\partial_\rho \equiv \partial_z$ on
Eq.~(\ref{Tave}). However, it is more instructive for what follows
to compute at least the first derivative {\em before} performing the
Gaussian average over $\rho_s$, that is, by first differentiating
the single--event profile (\ref{Tstep}) and then averaging. By using
 \be \partial_\rho T (\rho)\,=\,-\gamma\Theta(\rho - \rho_s)
 \,{\rm e}^{-\gamma(\rho - \rho_s)},\ee
which implies
 \be\label{aveder}
  \partial_\rho \lan T (\rho)\ran_Y\,=\,-\gamma
 \int \limits_{-\infty}^{\rho}
    \rmd\rho_s\, {\rm e}^{-\gamma(\rho - \rho_s)}
    \, P_Y(\rho_s)\,,\ee
it becomes manifest that the derivative is fully controlled by the
tail, as anticipated. But Eq.~(\ref{aveder}) also shows that the
contributions of the fronts with $\rho_s\ll \rho$ are exponentially
suppressed; so, unless $\rho$ is too different from $\lan \rho_s
\ran$ (in such a way that the probability $P_Y(\rho_s\sim\rho)$ be
suppressed even stronger !), the integral in Eq.~(\ref{aveder}) is
dominated by $\rho_s$ close to the upper limit, within a distance
$1/\gamma$ from it. This yields the following estimate
 \be\label{aveder1}
  \partial_\rho \lan T (\rho)\ran_Y\,\approx\,- P_Y(\rho_s =\rho)
  \,=\,-\frac{1}{\sqrt{\pi}\sigma}\,
   {\rm e}^{-\f{z^2}{\sigma^2}}\qquad{\rm for}\qquad
    |z|\ll \sigma^2\,,
  \ee
which can be confirmed via a direct calculation based on
Eq.~(\ref{Tave}). Remarkably, this result coincides with the first
derivative of the `black spots' approximation for $\lan T
(\rho)\ran_Y$, Eq.~(\ref{Thighsigma}).

\begin{figure}[t]
    \centerline{\epsfxsize=14cm\epsfbox{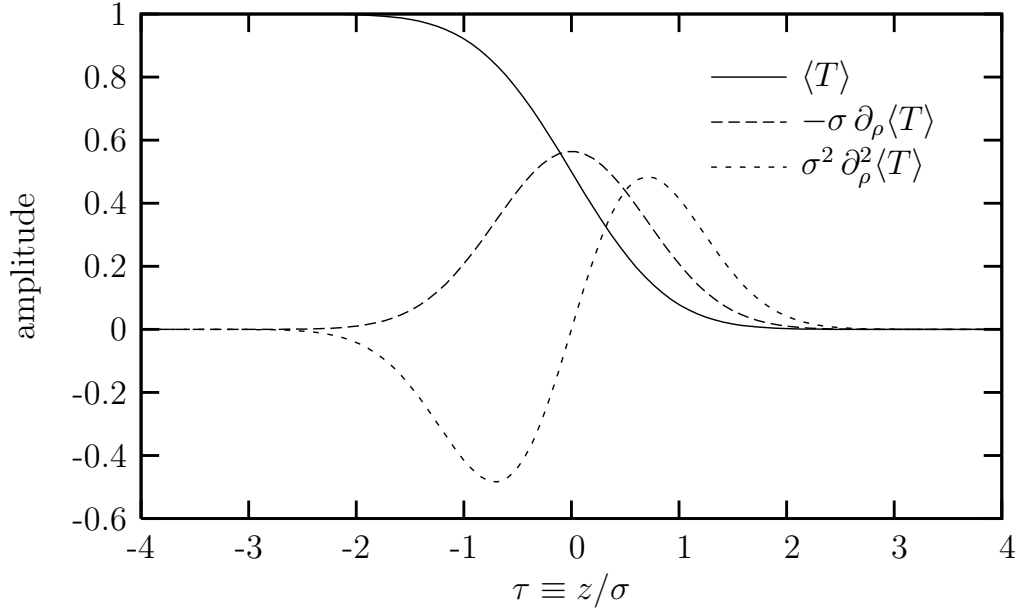}}
    \caption{\sl  The average dipole amplitude in the `black spots'
    approximation, Eq.~(\ref{Thighsigma}), together with its first and
    second derivatives (scaled by appropriate powers of $\sigma$ in such
    a way to exhibit diffusive scaling)
    are represented as functions of the respective scaling variable
    $\tau$, Eq.~(\ref{taudef}).
    \label{frontBS}}\bigskip
    \end{figure}

Thus, as anticipated, the $\rho$--derivative of the average dipole
amplitude is dominated by those fronts in the ensemble for which the
position $\rho_s$ of the front is of the order of the dipole
resolution scale. Because of that, the result (\ref{aveder1}) is
{\em robust and universal} (i.e., insensitive to the details of the
evolution), except possibly for its overall normalization, which is
not fully under control in the present approximations (since this
depends upon the precise approach of an individual front towards
saturation). The same conclusions hold, of course, for the
second--order derivative $\partial_\rho^2$, which can be computed
either from Eq.~(\ref{Tave}), or directly from
Eq.~(\ref{Thighsigma}), with the following result:
 \comment{ . Namely, one finds
 \be
\f{\partial^2}{\partial\rho^2}\lan T (\rho)\ran_Y \,=\,\g^2
 {\rm e}^{-z^2/\sigma^2}\left[
\lr{1-\f12\mbox{Erfc}\lr{\f{z}{\sigma}-\f{\g\sigma}2}}
 {\rm e}^{\lr{\f{z}{\sigma}-\f{\g\sigma}2}^2}-
 \f1{\sqrt{\pi}\g\sigma}\right]\
 ,\ee
 which for $|z|\ll \sigma^2$ simplifies to}
  \be\label{aveder2}
 \f{\partial^2}{\partial\rho^2}\lan T (\rho)\ran_Y \,\simeq\,
 \f{2\,z}{\sqrt{\pi}\sigma^3}\ {\rm e}^{-\f{z^2}{\sigma^2}} \qquad{\rm
 for}\qquad    |z|\ll \sigma^2\,. \ee
Since the estimates (\ref{aveder1}) and (\ref{aveder2}) are
independent of $\g$, it is clear that they identically hold for the
corresponding derivatives of $\lan T^2 (\rho)\ran_Y$ as well. This
is also in agreement with the fact that, within the validity range
of Eq.~(\ref{Thighsigma}), all the $N$--body dipole amplitudes
coincide with each other:
 \begin{equation}\label{TaveN}
   \lan T(\rho)\ran \,\simeq\,\,\langle T^2(\rho)\rangle \,\simeq\,
       \langle T^N (\rho)\rangle\,\approx\,
    \frac{1}{2}\, {\rm Erfc}\left(\frac{z}{\sigma} \right)
     \quad \mbox{for} \quad \sigma\gg 1
     \quad \mbox{and} \quad z \ll \sigma^2\,.
 \end{equation}
The universal functions which yield the dipole amplitude $\lan T
(\rho)\ran_Y$ and its first two derivatives (appropriately rescaled
by factors of $\sigma$ in order to exhibit diffusive scaling) in
this high--energy regime %($\sigma\gg 1$) and for $|z|\ll \sigma^2$
are shown in Fig. \ref{frontBS}.

%(In the case of $\lan T (\rho)\ran_Y$,
%the approximation (\ref{Thighsigma}) is valid
%down to arbitrarily large negative values of $z$.)

To complete our calculation of the generalized gluon distribution,
we need to evaluate the Fourier transform in Eq.~(\ref{Phirho}). We
shall first give an analytic estimate, based on Eq.~(\ref{aveder2}),
for the dominant behaviour at high energy, and then verify this
estimate, and also study a wider kinematical range, via a numerical
calculation using the more complete expression (\ref{Tave}) for the
average dipole amplitude.

Let us start by introducing some new notations, which will be useful
in what follows:
 \be\label{Zdef}
 Z \,\equiv\,\ln\frac{k^2}{\langle Q_s^2 \rangle}\,,\qquad
 \eta\,\equiv\,\ln\frac{1}{r^2k^2}\,=\,\ln\frac{1}{r^2\langle Q_s^2 \rangle}
 \,-\,\ln\frac{k^2}{\langle Q_s^2 \rangle}\,=\,z - Z\,,\ee
We shall now deduce an analytic estimate for the `occupation factor'
$\Phi(k,Y)\equiv \Phi(Z,Y)$ valid in high energy regime at
$\sigma\gg 1$ and for transverse momenta $k$ within a wide interval
around the average saturation momentum $\langle Q_s(Y) \rangle$,
specified by $|Z|\ll \sigma^2$. By inspection of the integrand in
Eq.~(\ref{Phirho}), we expect the integral to be dominated by dipole
sizes $r\sim 1/k$ (or $z\sim Z$), for which Eq.~(\ref{aveder2})
applies: Indeed, the Fourier transform restricts the integration to
$kr\simle 1$ (or $z\simge Z$), but the extremely small values $r\ll
1/k$ (or very large $z\gg Z$) are suppressed by the Gaussian decay
of the dipole amplitude, cf. Eq.~(\ref{aveder2}).

After also using the property $2\lan T \ran - \lan T^2 \ran = \lan T
\ran$, cf. Eq.~(\ref{TaveN}), and the new notations introduced in
Eq.~(\ref{Zdef}), Eq.~(\ref{Phirho}) can be rewritten as
 \be\label{Phirho1}
 \Phi(Z,Y)&\,=\, &2
 \int\rmd \eta\ J_0\lr{{\rm e}^{-\eta/2}}\,
 \f{\partial^2}{\partial \eta^2}\big\langle T (\eta+Z)
 \big\rangle_Y\,\ee
where the average amplitude is viewed as a function of $z=\eta+Z$
--- the natural variable in the r.h.s. of Eq.~(\ref{aveder2}).
The Bessel function $J_0(x)$ is oscillating for large values of its
argument $x$, with the amplitude of the oscillations rapidly
decreasing when increasing $x$. Hence, the dominant contribution to
the above integral is given by the interval between $x=0$ and the
first zero $x_0$ of the Bessel function (i.e., from $\eta=\eta_0$,
with ${\rm e}^{-\eta_0/2}\equiv x_0$, to $\eta\to \infty$), within
which $J_0(x)\approx $ const. This yields the following estimate
  \be\label{PhiBS}
 \Phi(Z,Y)&\,\approx\, & \f{1}{\sigma^3}
 \int\limits_{\eta_0}^\infty
 \rmd \eta\ (\eta+Z)\, {\rm e}^{-\f{(\eta+Z)^2}{\sigma^2}} \nn
 &\,=\, & \f{1}{\sigma}\, {\rm e}^{-\f{(\eta_0+Z)^2}{\sigma^2}}
 \,\approx\,\f{1}{\sigma}\, {\rm e}^{-\f{Z^2}{\sigma^2}}
  \qquad \mbox{for} \quad \sigma\gg 1
     \quad \mbox{and} \quad 1\ll |Z| \ll \sigma^2
 \,,\ee
up to some numerical fudge factor.  (The precise position of the
center of the Gaussian is not really under control here, hence the
restriction to $|Z|\gg 1$.)

Eq.~(\ref{PhiBS}) is essentially the same as the Gaussian
probability distribution of the fronts, Eq.~(\ref{probdens}). This
result can be physically understood as follows: $\Phi(Z,Y)$ with
$Z\equiv \ln({k^2}/{\langle Q_s^2 \rangle})$ is the {\em average}
`gluon occupation factor', as obtained after averaging the
corresponding {\em event--by--event} quantity with the Gaussian
probability (\ref{probdens}). In turn, the event--by--event
`occupation factor' $\Phi_{\rm event}(k,Y)$ is computed as the
appropriate Fourier transform, cf. Eq.~(\ref{Phidef}), of the dipole
amplitude $T$ in a single event\footnote{Strictly speaking, one
cannot use a piecewise interpolation, so like Eq.~(\ref{Tevent}), in
the calculation of a Fourier transform, since the discontinuities in
the higher derivatives would introduce spurious oscillations.
Rather, to that purpose one needs a smooth approximation to $T(r)$
so like the McLerran--Venugopalan model \cite{MV} or the BK
solution. Within such approximations, one finds indeed an
`occupation factor' $\Phi_{\rm event}(k,Y)$ with the behaviour
discussed in the text \cite{BraunG,Nestor03,KKT,BGV04}.},
Eq.~(\ref{Tevent}). The result of this calculation is well known
from previous studies based on the mean field approximation
% (which, as already discussed, is indeed similar to the
%event--by--event picture in the presence of fluctuations);
(see, e.g., Ref. \cite{KKT}) --- namely, one finds that the
single--event `occupation factor' vanishes as $k^2$ at small $k\ll
Q_s$ and it approaches the bremsstrahlung spectrum $\propto 1/k^2$
at large $k\gg Q_s$. In logarithmic units, with $\rho\equiv
\ln(k^2r_0^2)$, $\Phi_{\rm event}(\rho,Y)\sim {\rm
e}^{-|\rho-\rho_s|}$ is strongly peaked near $\rho=\rho_s$, and thus
it behaves almost like a $\delta$--function when averaged over
$\rho_s$ with a Gaussian probability with a large width $\sigma\gg
1$. This explains the result in Eq.~(\ref{PhiBS}). Note that the
latter ceases to apply when $|Z| \simge \sigma^2$, i.e., for momenta
$k$ which are very much smaller, or very much larger, than the
average saturation momentum; but in that case, we expect to recover
the standard exponential (in $\rho$) behaviour, namely $\Phi\sim
{\rm e}^{-|\rho-\lan\rho_s\ran|}$.

This discussion allows us to put the results in Sect.
\ref{SECT_DIPOLE} in a somewhat wider physical perspective: The
results of the high--energy evolution with Pomeron loops can be
visualized directly in terms of the (unintegrated) gluon
distribution in the hadron, independently of the problem of the
scattering. This perspective is consistent with the original
derivation of the Pomeron loop equations \cite{IT04,MSW05}, in which
the high--energy evolution was described as the color--glass
evolution of the target wavefunction.

\begin{figure}[t]
    \centerline{\epsfxsize=14cm\epsfbox{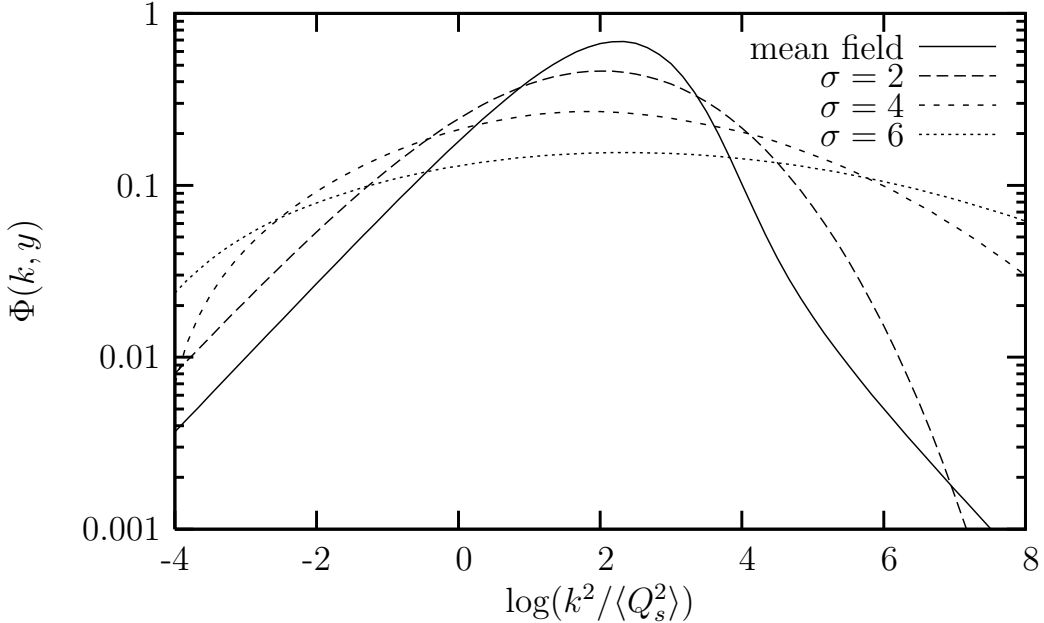}}
    \caption{\sl  The average `gluon occupation factor'
    obtained by numerical integration in Eq.~(\ref{Phirho})
    with the average dipole amplitude taken from Eq.~(\ref{Tave})
    (with $\gamma =1$) for $\sigma\ge 2$. For comparison, we also
    show the corresponding mean field curve obtained with the
    McLerran--Venugopalan model for the dipole amplitude.
    Note the flattening of the curve with increasing $\sigma$: this
    demonstrates the rapid breakdown of geometric scaling through
    fluctuations when increasing the energy.
    \label{gluon}}\bigskip
    \end{figure}

\begin{figure}[t]
    \centerline{\epsfxsize=14cm\epsfbox{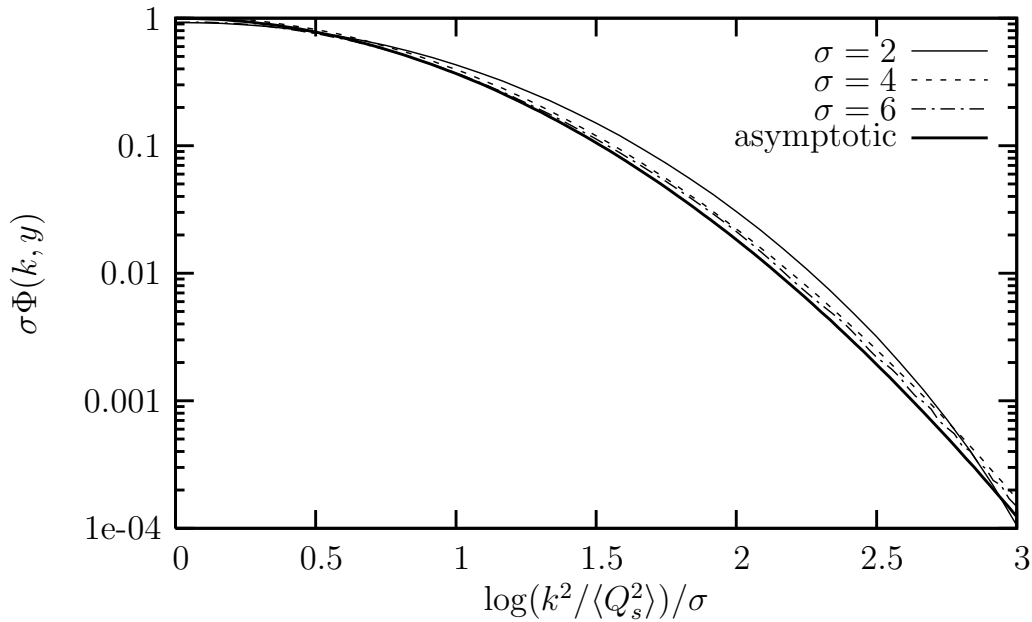}}
    \caption{\sl  The results corresponding to $\sigma\ge 2$
     in Fig. \ref{gluon} are redisplayed as a function of the diffusive
     scaling variable $\tau\equiv Z/\sigma$. One can thus appreciate
     the rapid onset of the scaling behaviour when increasing $\sigma$
     (or $Y$). The asymptotic curve is the Gaussian in the r.h.s. of
     Eq.~(\ref{PhiBS}).
    \label{gluonDS}}\bigskip
    \end{figure}

Note the contrast between the event--by--event `occupation factor'
$\Phi_{\rm event}(k,Y)$, which is strongly peaked at a transverse
momentum  $k$ of the order of the saturation scale $Q_s$ in that
event, and its expectation value $\Phi(k,Y)$, which at large $Y$ has
only a mild peak at $k\sim \langle Q_s(Y) \rangle$. This behaviour
is illustrated in Fig. \ref{gluon} which shows the quantity
$\Phi(Z,Y)$ obtained by numerical integration in Eq.~(\ref{Phirho})
with the average dipole amplitude taken from Eq.~(\ref{Tave}). As
manifest on this figure, the dispersion in the gluon distribution
increases rapidly with $\sigma$ (hence, with $Y$), leading to the
breakdown of geometric scaling. The large--$\sigma$ behaviour is in
qualitative agreement with the analytic estimate (\ref{PhiBS}). To
more precisely test the latter, notice that, according to
Eq.~(\ref{PhiBS}), the function $\sigma \Phi(Z,Y)$ should exhibit
diffusive scaling at sufficiently high energy. In order to check
this, we have redisplayed the numerical results from Fig.
\ref{gluon} as a function of the scaling variable $\tau\equiv
Z/\sigma$ : as manifest on Fig. \ref{gluonDS}, the diffusive scaling
emerges indeed (and quite fast !) when increasing $\sigma$.

\section{The cross--section for gluon production}
\setcounter{equation}{0}\label{SECT_SIGMA}

We are finally in a position to complete our original objective in
this paper, namely the calculation of the differential
cross--section (\ref{bfkl}) for forward gluon production in
onium--hadron scattering at high energy. By using the expression
(\ref{Tave}) for the average dipole--target amplitude, together with
the BFKL solution (\ref{tilden}) for the dipole number density in
the projectile, it is in principle possible to (numerically)
evaluate the integral in Eq.~(\ref{bfkl}) for arbitrary values of
its external variables. However, one can always gain more insight
via an analytic estimate, and this can be indeed obtained in the
kinematical regime of interest for us here, namely the regime in
which the Pomeron loop effects are most visible on this
cross--section.

Specifically, this regime is characterized by large values for the
total rapidity $Y$ and also for the pseudo--rapidity $\eta$ of the
produced gluon (but such that the difference $y_1=Y/2-\eta$ remains
relatively small, cf. Eq.~(\ref{Ycond})), and also by relatively
large values for the transverse momentum $k\equiv k_\perp$ of the
produced gluon, in the vicinity of the average saturation momentum
in the target $\langle Q_s^2(y_2) \rangle$. The most interesting
conditions are when $\sigma^2\equiv D_{\rm fr}\bar\alpha_s y_2 \gg
1$ and $|\ln(k^2/\langle Q_s^2 \rangle)|\ll \sigma^2$, since in
these conditions we probe the fluctuation--dominated regime in the
target wavefunction, cf. Fig. \ref{phase}. (Here and from now on,
the quantities $\langle Q_s^2 \rangle$ and $\sigma$ will always
refer to the target, hence they are evaluated at a rapidity
$y_2=Y/2+\eta$.)

Under these assumptions, the convolution in Eq.~(\ref{bfkl})
simplifies considerably, as is best seen on the
$k_\perp$--factorized version of this cross--section,
Eq.~(\ref{kTgluon}): Given the large asymmetry between the
evolutions of the projectile and respectively the target (recall
that $y_2\gg y_1$), the interesting values for the external momentum
$k^2$ are much larger than the momentum scale $Q_p^2(y_1)\sim
(1/r_0^2){\rm e}^{\lambda\abar y_1}$ which is generated in the
projectile wavefunction by its own high--energy evolution.
($Q_p^2(y_1)$ is the saturation scale of the projectile, and
effectively acts as an infrared cutoff\footnote{Indeed, this is the
scale at which there is a turnover in $\varphi({\bm p},y_1)$ from
the bremsstrahlung spectrum $\propto 1/p^2$ at large momenta to the
saturation spectrum $\propto p^2$ at low momenta.} in the integral
over $p$ in Eq.~(\ref{kTgluon}).) When this happens, the integral in
Eq.~(\ref{kTgluon}) is dominated by momenta $p$ logarithmically
distributed within the range $Q_p^2(y_1)\ll p^2 \ll k^2$; in the
leading logarithmic approximation w.r.t. the transverse logarithm
$\ln (k^2/Q_p^2)$, the result can be estimated as
 \be\label{collgluon}
 \f{\rmd\sigma}{\rmd\eta\rmd^2k  \rmd^2b}\,\approx\,
 \f{\abar}{k^2}\, \Phi(k,y_2)\int\limits^{\ k^2}
 \frac{\rmd^2 {\bm p}}{(2\pi)^2}\,\varphi({\bm p},y_1)
 \,\propto \,\f{1}{k^2}\, \Phi(k,y_2)\ x_1G_p(x_1,k^2)
  \,,\ee
where $x_1G_p(x_1,k^2)$ is the integrated gluon distribution in the
projectile evaluated for a longitudinal momentum fraction
$x_1=\ex{-y_1}$ and a transverse resolution (or momentum cutoff)
$k^2$. More precisely, it can be checked that the estimate
(\ref{collgluon}) holds under the following conditions:
%Let us give a brief derivation of this estimate which
%clarifies its validity range :
 \be\label{collcond}
 -\sigma\,\ll\,Z \equiv\,\ln\frac{k^2}
 {\langle Q_s^2 \rangle}\,
 \ll\,\sigma^2\,\qquad
 {\rm and}\qquad  \ln\frac{k^2}{Q_p^2}\,\gg\,\sigma\,
 .\ee
The first inequality ($Z\gg -\sigma$) is necessary to ensure that
the contribution of the relatively large momenta $p\gg k$ to the
integral in Eq.~(\ref{kTgluon}) is indeed subleading in the regime
where $\Phi(p,y_2)$ is given by Eq.~(\ref{PhiBS}). The second
inequality ($Z\ll \sigma^2$) ensures the validity of the
approximation (\ref{PhiBS}) for $\Phi(k,y_2)$. Finally, the third
inequality is not really independent
--- it automatically follows from the previous conditions together
with the fact that $\ln(\langle Q_s^2 \rangle/{Q_p^2})\propto \eta$
grows much faster with $\eta$ than $\sigma\sim\sqrt{\eta}$ --- but
it is nevertheless shown here to emphasize that the leading--twist
approximation for the projectile requires the stronger condition
$\ln({k^2}/{Q_p^2})\gg\sigma$ rather than just the usual one
$\ln({k^2}/{Q_p^2})\gg 1$.

Eq.~(\ref{collgluon}) is recognized as the {\em collinear
factorization} of the cross--section for gluon production; this is
valid, as expected, at sufficiently large $k$. Note that this
formula still involves the {\em unintegrated} gluon distribution of
the target, $\Phi(k,y_2)$, which is moreover computed in a regime
where the saturation effects are important. Thus, although it looks
familiar, the collinear factorization in Eq.~(\ref{collgluon})
emerges here in a less familiar context: the one in which the
leading--twist approximation applies to the projectile, but not also
to the target.

Strictly speaking, the projectile distribution $x_1G_p(x_1,k^2)$
which enters Eq.~(\ref{collgluon}) should be computed in the
double--logarithmic approximation (DLA) ---  the approximation which
resums both the high--energy logarithms $\ln(1/x_1)$ and the
transverse ones $\ln(k^2r_0^2)$ ---, since it emerges from a
$k_\perp$--factorization at high energy. But this is not important
here since, irrespective of the actual approximation which is used
to compute this function (DGLAP or DLA),  the result is anyway
slowly varying with $k^2$ as compared to the other functions which
appear in the r.h.s. of Eq.~(\ref{collgluon}).

According to Eq.~(\ref{collgluon}), the spectrum of the produced
gluon is proportional to the gluon occupation factor in the target
$\Phi(k,y_2)$ that we have already computed (in the high--energy
regime of interest) in the previous section. To better appreciate
the modifications introduced by Pomeron loop effects, it is
preferable to consider the distribution in $\ln k^2$ :
 \be\label{k2gluon}
 \f{\rmd\sigma}{\rmd\eta\,\rmd\! \ln\! k^2\,\rmd^2b}%\,\equiv\,
 \, \propto \, \Phi(k,y_2)\,,\ee
which is simply proportional to $\Phi(k,y_2)$. (We ignore the weak
dependence of the projectile gluon distribution upon $k^2$.) One can
then distinguish various physical regimes:

\texttt{i)} When $k^2$ is extremely high, so high that the
saturation effects become completely negligible, the gluon
occupation factor in the target takes its standard perturbative
form, i.e., the bremsstrahlung spectrum (up to the effects of the
DGLAP evolution, that we neglect here) :
 \be\label{highk2gluon}
 \f{\rmd\sigma}{\rmd\eta\,\rmd\! \ln\! k^2\,\rmd^2b}%\,\equiv\,
 \, \propto \, \f{1}{k^2}\qquad{\rm when}\qquad k^2 \ggg
 \langle Q_s^2 \rangle
 \,.\ee
We use here the notation $k^2 \ggg \langle Q_s^2 \rangle$ to
emphasize that, in the fluctuation--dominated regime at high energy
($\sigma^2\gg 1$), this leading--twist result is recovered only for
momenta so large as compared to $\langle Q_s^2 \rangle$ that the
condition  $Z \simge \sigma^2$ is satisfied, with $Z\equiv
\ln(k^2/\langle Q_s^2 \rangle)$.

For lower momenta, but still much larger than the (average)
saturation momentum, we encounter {\em saturation} effects, which
can be different depending upon the energy range at hand (see also
Fig. \ref{phase}) :

\texttt{ii)} For intermediate energies, such that $\sigma^2\ll 1$,
the effects of fluctuations are negligible and the BFKL evolution
with saturation boundary conditions leads to {\em geometric scaling}
at intermediate momenta, with an `anomalous dimension'
$\gamma_0\approx 0.63$ :
 \be\label{geomk2gluon}
 \f{\rmd\sigma}{\rmd\eta\,\rmd\! \ln\! k^2\,\rmd^2b}%\,\equiv\,
 \, \propto \, \left(\f{Q_s^2}{k^2}\right)^{\gamma_0}
 \qquad{\rm when}\qquad Q_s^2 \ll
 k^2 \simle Q_g^2
 \,.\ee
Here, we have written $\langle Q_s^2 \rangle\equiv Q_s^2$ (no
fluctuations !) and used $Q_g^2$ to denote the upper boundary of the
geometric scaling window. We have $Q_g^2\equiv Q_s^2 \ex{z_g}$ where
the exponent $z_g$ grows like $z_g \sim \sqrt{y_2}$ in the early
stages of the evolution (cf. Eq.~(\ref{Texp})), and then get stuck
at a value $z_g\sim L$ when $y_2\simge Y_{\rm form}$  (cf.
Eqs.~(\ref{L})--(\ref{Dtau})).

\texttt{iii)} For very high energies, such that $\sigma^2\gg 1$, one
enters the fluctuation--dominated regime, where the target
occupation factor $\Phi(k,y_2)$ (cf. Eq.~(\ref{PhiBS})), and hence
the produced gluon spectrum (\ref{k2gluon}), are dominated by `black
spots' with $Q_s\sim k$. Then, the average spectrum is a Gaussian in
$Z\equiv \ln(k^2/\langle Q_s^2 \rangle)$ and shows {\em diffusive
scaling} (after multiplication by a factor of $\sigma$) :
 \be\label{diffk2gluon}
 \f{\rmd\sigma}{\rmd\eta\,\rmd\! \ln\! k^2\,\rmd^2b}%\,\equiv\,
 \, \propto \, \f{1}{\sigma}\ {\exp}\left\{-\f{Z^2}{\sigma^2}\right\}
  \qquad \mbox{when} \qquad  -\sigma \ll Z
 \ll\sigma^2
 \,.\ee
Note the dramatic change in behaviour from the bremsstrahlung
spectrum $1/k^2 \propto \ex{-Z}$, which is naively expected at high
$k^2$, to the Gaussian spectrum (\ref{diffk2gluon}), which at
sufficiently high energy should prevail up to very large values of
$k^2$, well above $\langle Q_s^2 \rangle$ (and also well below it,
within the range indicated in the equation above).

\section{Conclusion and perspectives}
\setcounter{equation}{0} \label{SECT_CONC}
%\newpage

The main conclusion of this work is that forward particle production
in proton--proton collisions at high energy represents an ideal
laboratory to search for the new physics expected in QCD at very
high energies. Indeed, the spectrum of the gluons produced at
forward rapidities reflects quite faithfully the unintegrated gluon
distribution in the highly evolved hadronic target, and thus is
directly sensitive to new phenomena like gluon saturation and
gluon--number fluctuations, which are expected to become important
at high energy. The essential consequence of these phenomena is the
fact that, up to relatively large transverse momenta --- well above
the {\em average} saturation momentum in the target wavefunction
---, the cross--section for gluon production is dominated by {\em
black spots}, i.e., rare gluon configurations with unusually large
density, which are at saturation on the resolution scale of the
produced gluon. And the distinctive signature of this behaviour,
which should be its hallmark in the experimental results, is a new
scaling law --- the {\em diffusive scaling} ---, which should
eventually replace at sufficiently high energy the `geometric
scaling' behaviour currently observed in the small--$x$ data at HERA
and RHIC.

Both geometric scaling and diffusive scaling are exceptional
phenomena, in that they represent consequences of saturation which
manifest themselves at relatively large transverse momenta, well
outside the saturation region --- that is, in a region of the
phase--space where one would naively expect the applicability of the
linear, or `leading--twist', evolution equations of perturbative QCD
(DGLAP or BFKL). But although they are both hallmarks of saturation,
these two types of scaling correspond to very different physical
regimes, that they should unambiguously identify in the data :

\begin{itemize}

\item {\em Geometric scaling} occurs at {\em intermediate
energies}, i.e., energies which are large enough for the saturation
momentum $Q_s$ to be a hard scale, yet low enough for the dispersion
in the values of $Q_s$ (as generated via gluon--number fluctuations
at low density) to remain negligible ($\sigma\ll 1$, cf. Fig.
\ref{phase}). In that regime, `geometric scaling' is simply the
statement that the physics remains invariant along any line parallel
to the saturation line $\rho_s(Y)=\lambda Y$. But the dynamics when
increasing $\rho\equiv\ln k_\perp$ above $\rho_s$ (i.e., when moving
away from saturation) is still described by the leading--twist,
BFKL, formalism.

\vskip 1mm
\item {\em Diffusive scaling}, on the other hand, represents
the ultimate behaviour at {\em sufficiently high energy}, where the
dispersion in the gluon configurations is very large ($\sigma\gg 1$)
and the high--density fluctuations dominate the expectation values
within a wide window around the {\em average} saturation line $\lan
\rho_s(Y)\ran$ --- namely, so long as $|\rho-\lan \rho_s\ran|\ll
\sigma^2$. Within that whole window (which includes very large
momenta $k_\perp\gg  \lan Q_s\ran$ !), the leading--twist
approximation breaks down and the cross--section for gluon
production is controlled by the physics of saturation (similarly to
the DIS cross--sections \cite{HIMST06}).

\end{itemize}

Properties of the gluon distribution produced via a high--energy
evolution, the geometric and diffusive scaling naturally transmit
(within their respective ranges of existence) to the scattering
amplitudes for external dipoles --- the most direct probes of the
gluon distribution in the target --- and hence to all the processes,
so like lepton--hadron DIS or {\em forward} gluon production in
hadron--hadron collisions, which can be given a dipole
factorization. On the other hand, the consequences of these
phenomena on the particle production at {\em central} rapidities
--- and, more generally, on the {\em symmetric} collision between
two highly--evolved hadrons --- are presently unclear, since there
is no general factorization scheme for the scattering between two
hadronic systems which are both affected by saturation.

This problem becomes particularly acute in the high--energy regime
where the Pomeron loop effects are important, because in that case
we expect saturated gluon configurations to significantly contribute
to scattering up to very large transverse momenta (at least, this is
what we have seen to happen in DIS and for the forward gluon
production), so the standard, collinear-- or $k_\perp$--,
factorization schemes fail to apply even at very large transverse
momenta, well above the average saturation scale. It thus remains as
an important open problem how to compute, e.g., gluon production at
central rapidity in proton--proton scattering with Pomeron loops. It
is tempting to conjecture that, within a wide kinematical range,
this process should be dominated by black--spots--on--black--spots
scattering, but this remains to be proven.

But further theoretical progress is also needed in the simpler case
of the asymmetric, dilute--dense, collisions, as discussed in this
paper and in Ref. \cite{HIMST06}. The results presented here have
been obtained via rather crude approximations, which rely on the
correspondence between high--energy evolution in QCD and the
reaction--diffusion process in statistical physics, but do not allow
for a more detailed characterization of the QCD amplitudes. Although
the qualitative conclusions which emerged in this way (like the
dominance of black spots up to large transverse momenta or the
diffusive scaling) are quite robust, as they follow from fundamental
considerations, some other, equally important, properties ---
chiefly among them, the energy dependencies of the average
saturation momentum and of its dispersion ---, remain poorly
understood and deserve further studies.

To make progress, one needs more complete (possibly numerical)
solutions to the Pomeron loop equations of Ref. \cite{IT04}; this
would allow one not only to determine the parameters $\lambda$ and
$D_{\rm fr}$, but also to control the impact--parameter dependence
of the amplitudes (to leading order accuracy in $\alpha_s$ and
$1/N_c$). Furthermore, the previous experience with the
next--to--leading--order (NLO) corrections in the context of
saturation \cite{DT02} shows that such corrections are
quantitatively important for all but the asymptotically high
energies. A NLO formalism including Pomeron loops is therefore
highly desirable, but this looks quite difficult to achieve.
(Already the generalization of the present, LO, equations to generic
values of $N_c$ meets with serious, technical and conceptual,
problems, in spite of some recent progress
\cite{KL05,BREM,Balit05}.) There should be nevertheless possible to
estimate at least {\em some} NLO effects, like those of the running
of the coupling or of imposing energy conservation in the evolution.
Last but not least, in view of more realistic calculations of the
(forward) hadronic yield, it would be interesting to include some
more processes at partonic level, so like quark pair production, and
also to complete the (generalized) collinear factorization,
Eq.~(\ref{collgluon}), with fragmentation functions for the produced
partons.

\section*{Acknowledgments}

We are grateful to L. McLerran for encouraging us to write down a
pedagogical account of the `black spots' picture at high energy. We
acknowledge useful conversations with F. Gelis, K. Itakura,
R.~Peschanski, and D. N. Triantafyllopoulos. We would like to thank
the Theory Group at KEK (Tsukuba, Japan) for hospitality during the
gestation of this work. G.S. is funded by the National Funds for
Scientific Research (Belgium).

%\bibliographystyle{unsrt}
%\bibliography{myrefs}

\begin{thebibliography}{10}

\bibitem{RHIC-dAu-mid}
B.~B.~Back {\it et al.} [PHOBOS Collaboration], {\it Phys.\ Rev.\
Lett.}\ {\bf
  91} (2003) 072302; S.~S.~Adler {\it et al.} [PHENIX Collaboration], {\it
  Phys.\ Rev.\ Lett.}\ {\bf 91} (2003) 072303; J.~Adams {\it et al.} [STAR
  Collaboration], {\it Phys.\ Rev.\ Lett.}\ {\bf 91} (2003) 072304; I.~Arsene
  {\it et al.} [BRAHMS Collaboration], {\it Phys.\ Rev.\ Lett.}\ {\bf 91}
  (2003) 072305.

\bibitem{Brahms-data}
I.~Arsene {\it et al.} [BRAHMS Collaboration], {\it Phys.\ Rev.\
Lett.}\ {\bf
  93} (2004) 242303.

\bibitem{STAR-data}
J.~Adams {\it et al.} [STAR Collaboration], {\it ``Forward neutral
pion
  production in p+p and d+Au collisions at $\sqrt{s_NN}=200$ GeV''},
  arXiv:nucl-ex/0602011.

\bibitem{KLM02}
D. E. Kharzeev, E. Levin, and L. McLerran, {\it Phys. Lett.} {\bf
B561} (2003)
  93.

\bibitem{JNV}
J.~Jalilian-Marian, Y.~Nara and R.~Venugopalan, {\it Phys.\ Lett.}\
{\bf B577}
  (2003) 54.

\bibitem{KKT}
D. Kharzeev, Yu. V. Kovchegov, and K. Tuchin, {\it Phys. Rev.} {\bf
D66} (2003)
  094013.

\bibitem{Baier03}
R.~Baier, A.~Kovner and U.~A.~Wiedemann, Phys.\ Rev.\ D {\bf 68}
(2003) 054009.

\bibitem{Nestor03}
J.L.~Albacete, N.~Armesto, A.~Kovner, C.A.~Salgado and
U.A.~Wiedemann, {\it
  Phys. Rev. Lett.} {\bf 92} (2004) 082001. % [arXiv:hep-ph/0307179].

\bibitem{BGV04}
J.P. Blaizot, F. Gelis, R. Venugopalan, Nucl.\ Phys.\ A {\bf 743}
(2004) 13 ;
  {\it ibid.} (2004) 57.

\bibitem{IIT04}
E.~Iancu, K.~Itakura, and D.N.~Triantafyllopoulos, {\it Nucl. Phys.}
{\bf A742}
  (2004) 182.

\bibitem{KKT2}
D. Kharzeev, Yu. V. Kovchegov, and K. Tuchin, {\it Phys.\ Lett.}
{\bf B599}
  (2004) 23.

\bibitem{BMTS06}
R. Baier, Y. Mehtar-Tani and D. Schiff, Nucl. Phys. {\bf A764}
(2006) 515.

\bibitem{Dumitru1}
A. Dumitru, A. Hayashigaki, and J. Jalilian-Marian, {\it Nucl.
Phys.} {\bf
  A765} (2006) 464.

\bibitem{Dumitru2}
A. Dumitru, A. Hayashigaki, and J. Jalilian-Marian, {\it Nucl.
Phys.} {\bf
  A770} (2006) 57.

\bibitem{IIM03}
E.~Iancu, K.~Itakura and S.~Munier, {\it Phys. Lett.} {\bf B590}
(2004) 199.

\bibitem{MV}
L.~McLerran and R.~Venugopalan, {\it Phys.\ Rev.}\ {\bf D49} (1994)
2233; {\it
  ibid.} {\bf 49} (1994) 3352; {\it ibid.} {\bf 50} (1994) 2225.

\bibitem{CGC}
E.~Iancu, A.~Leonidov and L.~McLerran, {\it Nucl. Phys.}~{\bf A692}
(2001) 583;
  {\it Phys. Lett.} {\bf B510} (2001) 133; E.~Ferreiro, E.~Iancu,
  A.~Leonidov,
  L.~McLerran, {\it Nucl. Phys.} {\bf A703} (2002) 489.

\bibitem{EdiCGC}
E.~Iancu, A.~Leonidov and L.~McLerran, {\it ``The Colour Glass
Condensate: An
  Introduction''}, arXiv:hep-ph/0202270. Published in {\it QCD Perspectives on
  Hot and Dense Matter}, Eds. J.-P.~Blaizot and E.~Iancu, NATO Science Series,
  Kluwer, 2002;\\ E.~Iancu and R.~Venugopalan, {\it ``The Color Glass
  Condensate and High Energy Scattering in QCD''}, arXiv:hep-ph/0303204.
  Published in {\it Quark-Gluon Plasma 3}, Eds. R.C.~Hwa and X.-N.~Wang, World
  Scientific, 2003.

\bibitem{Brahms-Review}
I.~Arsene {\it et al.} [BRAHMS Collaboration], {\it Nucl. Phys.}
{\bf A757}
  (2005) 1.

\bibitem{geometric}
A.M.~Stasto, K.~Golec-Biernat and J.~Kwiecinski, Phys. Rev. Lett.
{\bf 86}
  (2001) 596.

\bibitem{SCALING}
E.~Iancu, K.~Itakura, and L.~McLerran, {\it Nucl. Phys.} {\bf A708}
(2002) 327.

\bibitem{MT02}
A.H.~Mueller and D.N.~Triantafyllopoulos, {\it Nucl. Phys.} {\bf
B640} (2002)
  331.

\bibitem{DT02}
D.N.~Triantafyllopoulos, {\it Nucl. Phys.} {\bf B648} (2003) 293.

\bibitem{KM98}
Yu. V. Kovchegov and A. H. Mueller, {\it Nucl. Phys.} {\bf B529}
(1998) 451.

\bibitem{KTS99}
B.Z. Kopeliovich, A.V. Tarasov, and A. Schafer, {\it Phys. Rev.}
{\bf C59}
  (1999) 1609.

\bibitem{DM01}
A. Dumitru and L. McLerran, {\it Nucl. Phys.} {\bf A700} (2002) 492.

\bibitem{KT02}
Yu. V. Kovchegov and K. Tuchin, {\it Phys. Rev.} {\bf D65} (2002)
074026.

\bibitem{KW02}
A. Kovner and U. Wiedemann, {\it Phys. Rev.} {\bf D64} (2001)
114002.

\bibitem{CM04}
C. Marquet, {\it Nucl. Phys.} {\bf B705} (2005) 319.

\bibitem{BraunG}
M.~Braun, {\it Eur. Phys. J.} {\bf C16} (2000) 337; {\it Phys.
Lett.} {\bf
  B483} (2000) 105.

\bibitem{B}
I.~Balitsky, {\it Nucl.\ Phys.}\ {\bf B463} (1996) 99; {\it Phys.
Lett.} {\bf B518} (2001) 235; hep-ph/0101042.

\bibitem{JKLW}
J.~Jalilian-Marian, A.~Kovner, A.~Leonidov and H.~Weigert, {\it
Nucl.\ Phys.}\
  {\bf B504} (1997) 415; {\it Phys.\ Rev.}\ {\bf D59} (1999) 014014;
  J.~Jalilian-Marian, A.~Kovner, H.~Weigert, {\it Phys.\ Rev.}\ {\bf D59}
  (1999) 014015; A. Kovner, J. G. Milhano and H. Weigert, {\it Phys. Rev.} {\bf
  D62} (2000) 114005.

\bibitem{W}
H.~Weigert, {\it Nucl. Phys.} {\bf A703} (2002) 823.

\bibitem{K}
Yu.V.~Kovchegov, {\it Phys. Rev.} {\bf D60} (1999), 034008; {\it
ibid.} {\bf
  D61} (1999), 074018.

\bibitem{IM03}
E.~Iancu and A.H.~Mueller, {\it Nucl.\ Phys.}\ {\bf A730} (2004)
460; {\it
  Nucl.\ Phys.}\ {\bf A730} (2004) 494.

\bibitem{MS04}
A.H.~Mueller and A.I.~Shoshi, {\it Nucl.\ Phys.}\ {\bf B692} (2004)
175.

\bibitem{IMM04}
E.~Iancu, A.H.~Mueller and S.~Munier, {\it Phys.~Lett.~}{\bf B606}
(2005) 342.

\bibitem{IT04}
E.~Iancu and D.N.~Triantafyllopoulos, {\it Nucl.~Phys.~}{\bf A756}
(2005) 419;
  {\it Phys.~Lett.~}{\bf B610} (2005) 253.

\bibitem{MSW05}
A.H.~Mueller, A.I.~Shoshi, S.M.H.~Wong, {\it Nucl.~Phys.~}{\bf B715}
(2005)
  440.

\bibitem{LL05}
E.~Levin, M.~Lublinsky, {\it Nucl.~Phys.~}{\bf A763} (2005) 172.

\bibitem{KL05}
A.~Kovner, M.~Lublinsky, {\it Phys.~Rev.~}{\bf D71} (2005) 085004;
{\it
  Phys.~Rev.~Lett.~}{\bf 94} (2005) 181603.

\bibitem{BIIT05}
J.-P.~Blaizot, E.~Iancu, K.~Itakura, D.N.~Triantafyllopoulos, {\it
  Phys.~Lett.~}{\bf B615} (2005) 221.

\bibitem{BREM}
Y. Hatta, E. Iancu, L. McLerran, A. Stasto, and
D.N.~Triantafyllopoulos, {\it
  Nucl. Phys.} {\bf A764} (2006) 423, arXiv:hep-ph/0504182.

\bibitem{Balit05}
I. Balitsky, {\it Phys. Rev.} {\bf D72} (2005) 074027,
arXiv:hep-ph/0507237.

\bibitem{HIMST06}
Y.~Hatta, E.~Iancu, C. Marquet, G. Soyez, D.N.~Triantafyllopoulos,
{\it
  ``Diffusive scaling and the high--energy limit of deep inelastic scattering
  in QCD at large $N_c$''}, hep-ph/0601150.

\bibitem{AM94}
A.H.~Mueller, {\it Nucl. Phys.} {\bf B415} (1994) 373; A.H. Mueller,
B. Patel, {\it Nucl. Phys.} {\bf B425} (1994) 471; A.H.~Mueller,
{\it Nucl. Phys.} {\bf B437} (1995) 107.

\bibitem{Salam95}
G.P.~Salam, {\it Nucl. Phys.} {\bf B449} (1995) 589; {\it Nucl.
Phys.} {\bf
  B461} (1996) 512; A.H.~Mueller, G.P.~Salam, {\it Nucl. Phys.} {\bf B475}
  (1996) 293.

\bibitem{BFKL}
L.N.~Lipatov, {\it Sov.\ J.\ Nucl.\ Phys.}\,{\bf 23} (1976) 338;\\
E.A.~Kuraev,
  L.N.~Lipatov and V.S.~Fadin, {\it Zh. Eksp. Teor. Fiz} {\bf 72}, 3 (1977)
  ({\it Sov. Phys. JETP }{\bf 45} (1977) 199);\\ Ya.Ya.~Balitsky and
  L.N.~Lipatov, {\it Sov.\ J.\ Nucl.\ Phys.} {\bf 28} (1978) 822.

\bibitem{MP03}
S.~Munier and R.~Peschanski, {\it Phys. Rev. Lett.} {\bf 91} (2003)
232001;
  {\it Phys.\ Rev.}\ {\bf D69} (2004) 034008; {\it ibid.} {\bf D70} (2004)
  077503.


\bibitem{GS05}
G. Soyez, {\it Phys. Rev.} {\bf D72} (2005) 016007.

\bibitem{EGBM05}
R. Enberg, K. Golec--Biernat and S. Munier, {\it Phys. Rev.} {\bf
D72} (2005) 074021.

\bibitem{Saar}
For a recent review, see W.~Van Saarloos, {\it Phys. Rep.} {\bf 386}
(2003) 29.

\bibitem{BD97}
E.~Brunet and B.~Derrida, {\it Phys. Rev.} {\bf E56} (1997) 2597;
{\it Comp.
  Phys. Comm.} {\bf 121-122} (1999) 376; {\it J. Stat. Phys.} {\bf 103} (2001)
  269.

\bibitem{Kov05}
Yu.V. Kovchegov, {\it Phys.~Rev.~} {\bf D72} 094009,
arXiv:hep-ph/0508276.

\bibitem{Motyka}
K.~Golec-Biernat, L.~Motyka, and A.M.~Sta\'sto, {\it Phys. Rev.}
{\bf D65}
  (2002) 074037.

\bibitem{LT99}
E.~Levin and K.~Tuchin, {\it Nucl. Phys.} {\bf B573} (2000) 833;
{\it Nucl.
  Phys.} {\bf A691} (2001) 779; {\it Nucl. Phys. } {\bf A693} (2001) 787.

\bibitem{LL01}
E.~Levin and M.~Lublinsky, {\it Phys. Lett.} {\bf B521} (2001) 233;
{\it Eur.
  Phys. J.} {\bf C22} (2002) 647; M. Lublinsky, {\it Eur. Phys. J.} {\bf C21}
  (2001) 513.

\bibitem{RW03}
K.~Rummukainen and H.~Weigert, {\it Nucl.\ Phys.}\ {\bf A739} (2004)
183.

\bibitem{MS05}
C. Marquet, R. Peschanski and G. Soyez, {\it Nucl. Phys.} {\bf A756}
(2005)
  399; C. Marquet and G. Soyez, {\it Nucl. Phys.} {\bf A760} (2005) 208.

\bibitem{GLR}
L.V.~Gribov, E.M.~Levin, and M.G.~Ryskin, {\it Phys. Rept. } {\bf
100} (1983)
  1.

\bibitem{MPS05}
C. Marquet, R. Peschanski and G. Soyez, {\it ``Consequences of
strong fluctuations
  on high-energy QCD evolution''}, hep-ph/0512186.

\bibitem{BDMM}
E. Brunet, B. Derrida, A. H. Mueller and S. Munier, {\it ``A
phenomenological theory
  giving the full statistics of the position of fluctuating pulled
  fronts''},
  arXiv:cond-mat/0512021.

\end{thebibliography}

\end{document}